\documentclass[apjl, iop]{emulateapj}

\slugcomment{Accepted for publication in ApJ}
\usepackage[tight]{subfigure}
\usepackage[colorlinks]{hyperref}
\hypersetup{ colorlinks, linkcolor=blue, citecolor=blue, anchorcolor=blue }
\usepackage{color}
\usepackage{amsmath}
\newcommand{\B}[1]{{\color{blue}  #1}}

\graphicspath{{images/}}

\newcommand{\Msun}{${\rm M}_{\odot}$}
\newcommand{\kms}{km\,s$^{-1}$}
\newcommand{\SiII}{Si~{\sc ii}}

\newcommand{\NaII}{Na~{\sc ii}}
\newcommand{\NaI}{Na~{\sc i}~D}

\newcommand{\OI}{O~{\sc i}}
\newcommand{\OII}{O~{\sc ii}}
\newcommand{\CI}{C~{\sc i}}
\newcommand{\CII}{C~{\sc ii}}

\newcommand{\HeI}{He~{\sc i}}

\newcommand{\FeII}{Fe~{\sc ii}}
\newcommand{\CaII}{Ca~{\sc ii}}
\newcommand{\Nifs}{$^{56}$Ni}
\newcommand{\ld}{$\lambda$}

\shorttitle{SN\,2014L}
\shortauthors{Zhang et al.}

\begin{document}

\title{Optical Observations of the Young Type I\MakeLowercase{c} SUPERNOVA SN\,2014L in M99}
\author{Jujia Zhang\altaffilmark{1,2,3,4}, Xiaofeng Wang\altaffilmark{5}, J\'{o}zsef Vink\'{o}\altaffilmark{6,7,8}, J. Craig Wheeler\altaffilmark{8}, Liang Chang\altaffilmark{1,3,4}, Yi Yang\altaffilmark{9$\dagger$,2}, Lifan Wang\altaffilmark{2}, Qian Zhai\altaffilmark{1,3,4}, Liming Rui\altaffilmark{5}, Jun Mo\altaffilmark{5}, Tianmeng Zhang\altaffilmark{10}, Yu Zhang\altaffilmark{11}, Jianguo Wang\altaffilmark{1,3,4}, Jirong Mao\altaffilmark{1,3,4}, Chuanjun Wang\altaffilmark{1,3,4}, Weimin Yi\altaffilmark{1,3,4}, Yuxin Xin\altaffilmark{1,3,4}, Wenxiong Li\altaffilmark{5}, Baoli Lun\altaffilmark{1,3,4}, Kaixing Lu\altaffilmark{1,3,4}, Hanna Sai\altaffilmark{5}, Xiangming Zheng\altaffilmark{1,3,4}, Xiliang Zhang\altaffilmark{1,3,4}, Xu Zhou\altaffilmark{10}, and Jinming Bai\altaffilmark{1,3,4}}
\altaffiltext{1}{Yunnan Observatories (YNAO), Chinese Academy of Sciences, Kunming 650216, China; jujia@ynao.ac.cn.}
\altaffiltext{2}{George P. and Cynthia Woods Mitchell Institute for Fundamental Physics $\&$ Astronomy, Texas A. $\&$ M. University, College Station, TX 77843, USA.}
\altaffiltext{3}{Key Laboratory for the Structure and Evolution of Celestial Objects, Chinese Academy of Sciences, Kunming 650216, China.}
\altaffiltext{4}{Center for Astronomical Mega-Science, Chinese Academy of Sciences, 20A Datun Road, Chaoyang District, Beijing, 100012, China.}
\altaffiltext{5}{Physics Department and Tsinghua Center for Astrophysics (THCA), Tsinghua University, Beijing 100084, China.}
\altaffiltext{6}{Konkoly Observatory, Research Centre for Astronomy and Earth Sciences, Konkoly Thege ut 15-17, Budapest, 1121, Hungary}
\altaffiltext{7}{Department of Optics and Quantum Electronics, University of Szeged, H-6720 $Szeged$, D\'{o}mt\'{e}r 9., Hungary. }
\altaffiltext{8}{Department of Astronomy, University of Texas at Austin, Austin, TX 78712-1205, USA.}
\altaffiltext{9}{Department of Particle Physics and Astrophysics, Weizmann Institute of Science, Rehovot 76100, Israel}
\altaffiltext{10}{National Astronomical Observatories of China (NAOC), Chinese Academy of Sciences, Beijing 100012, China.}
\altaffiltext{11}{Xinjiang Astronomical Observatory(XAO), Chinese Academy of Sciences, Urumqi, 830011, China.}
%\altaffiltext{}{George P. and Cynthia Woods Mitchell Institute for Fundamental Physics \& Astronomy, Texas A. \& M. University, College Station, TX 77843, USA.}
\altaffiltext{$\dagger$}{Benoziyo Fellow} 

\begin{abstract}

We present optical spectroscopic and photometric observations of the nearby type Ic supernova (SN Ic) SN\,2014L. This SN was discovered by the Tsinghua-NAOC Transient Survey (TNTS) in the nearby type-Sc spiral galaxy M99 (NGC 4254). Fitting to the early-time light curve indicates that SN\,2014L was detected at only a few hours after the shock breakout, and it reached a peak brightness of  $M_{\rm V}\,=\,-17.73\,\pm\,0.28$ mag ($L\,=\,[2.06\,\pm\,0.50]\,\times\,10^{42}$ erg s$^{-1}$) approximately 13 days later. SN\,2014L shows a close resemblance to SN\,2007gr in the photometric evolution, while it shows stronger absorption features of intermediate-mass elements (especially \CaII) in the early-time spectra. Based on simple modeling of the observed light curves, we derived the mass of synthesized $^{56}$Ni as $M_{\rm Ni}$ =\,0.075\,$\pm$\,0.025\,M$_{\sun}$, and the mass and total energy of the ejecta as $M_{\rm ej}$=1.00$\pm$0.20\Msun\ and $E_{\rm ej}=1.45\pm0.25$ foe, respectively. Given these typical explosion parameters, the early detection, and the extensive observations,  we suggest that SN\,2014L could be a template sample for the investigation of SNe Ic.

\end{abstract}


\section{Introduction}
\label{sect:Intro}
Type Ic supernovae are characterized by the absence of hydrogen and helium in the spectra (see, e.g., reviews by \citealp{Filippenko97,Modjaz14,liu16,BW17}). The primary spectral distinction between regular Ic and Ib is the strength and evolution of helium lines. Recent studies show, however, that weak helium absorptions can be detected in some SNe Ic \citep{Chen14,Milisavljevic15}, while there are also arguments that the helium features are not present in SNe Ic (i.e., \citealp{liu16}). Theoretically, the appearance of helium in the spectra may not necessarily suggest a classification of an SN Ib, which might relate to the abundance ratio of nickel and helium in the outer layers.\citep{Wheeler87,Shigeyama90,Hachisu91}.

%%%%%%%%%%%%%%%%%%%%%%%
\begin{figure*}
\centering
\includegraphics[width=17cm,angle=0]{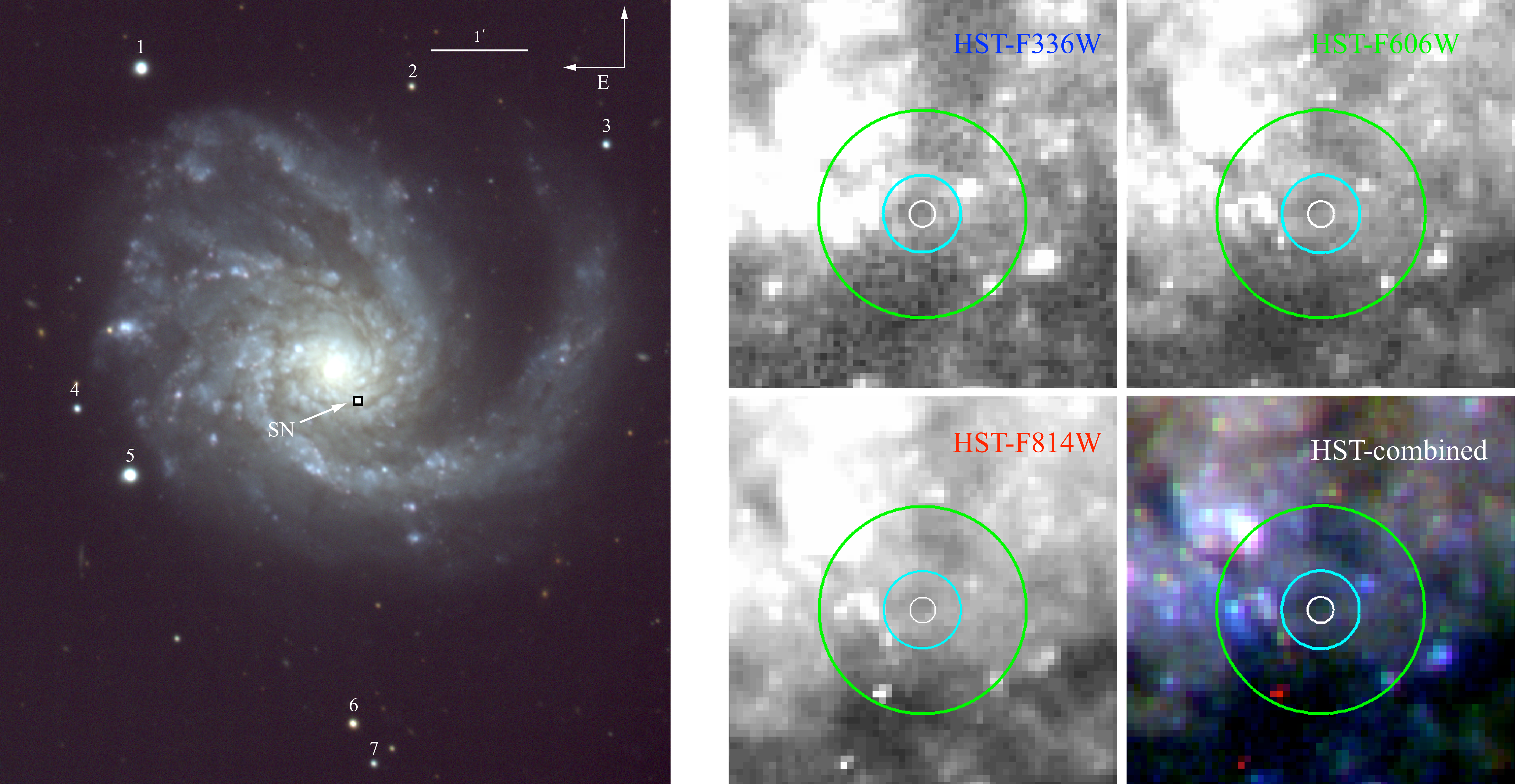}
 \caption{Left panel: the finder chart of SN\,2014L and its local reference, taken by the LJT and YFOSC on 2014 January. The mean FWHM of this combined image is $1\arcsec.60$ under the scale of $0\arcsec.28$/pixel. Right panel: the pre-explosion image, corresponding  to the black box on the left panel and including the birth place region of SN\,2014L, taken by $Hubble Space telescope$  Wide-Field Planetary Camera 2 in 2009 in the filters of F336W, F606W, and F814W under the scale of $0\arcsec.1$/pixel. The combined image is also shown. The circles having radii of 0$\arcsec$.2, 0$\arcsec$.6 and 1$\arcsec$.6, respectively, are centered at the location of SN.}
\label{<img>}
\end{figure*}
%%%%%%%%%%%%%%%%%%%%%%%

Observationally, SNe Ic are found to show large varieties in spectral features and line profiles. For example, the subclass of broad-lined SNe Ic (BL-Ic) is characterized by broad, highly blueshifted line profiles in their spectra, relatively high luminosity, and considerable kinetic energies ($2-5\times 10^{52}$ ergs, \citealp{Woosley06}). Some BL SNe Ic are found to be associated with long-duration gamma-ray bursts (GRBs), e.g., SN\,1998bw \citep{Galama98,Iwamoto98} and SN\,2003dh \citep{Hjorth03,Mazzali03}, and their progenitors are reported to have relatively low metallicity (e.g., \citealp{Woosley93,Woosley06}). Some peculiar SNe Ic events with prominent calcium features and fast spectral evolution, i.e., SN\,2012hn \citep{Valenti14}, which are dubbed as Ca-rich SNe Ic. This observed diversity of SNe Ic indicates that they may arise in different channels, e.g., single massive stars or binary system.

Photometric and spectroscopic observations have been published for dozens of SNe Ic \citep{Bianco14, Modjaz14}. However, the sample with very early observations and good phase coverage (e.g., SN\,1994I, \citealp{Wheeler94,Filippenko95,Richmond96}; SN\,2007gr, \citealp{Valenti08,Hunter09,Chen14}; SN\,2013ge, \citealp{Drout16}) is still limited. In this paper, we present  extensive photometric and spectroscopic observations  of the nearby SN Ic SN\,2014L, spanning from $\sim$10 days to $\sim$140 days relative to the maximum light.  This SN was detected at only a few hours after the shock breakout and might be one of the youngest SN Ic ever discovered. 

This paper is organized as follows. Observation and data reductions are described in Section \ref{sect:Obs}. The results from photometry and spectroscopy are presented in Section \ref{sect:LC} and \ref{sect:SP}, respectively. Explosion parameters are calculated and discussed in Section \ref{sect:Para}. A summary is given in Section \ref{sect:con}.

\section{Observations}
\label{sect:Obs}

SN\,2014L was discovered by the Tsinghua - NAOC (National astronomical observatories of China) transient survey (TNTS) on 2014 January 26.83 2014 UT  \citep{TMZhang14} in the nearby Sc-type galaxy M99 (= NGC 4254), at an unfiltered magnitude of 17.2. This survey uses a 0.6-m Schmidt telescope (+4K$\times$4K CCD) located at Xinglong Observatory in China \citep{TMZhang15}. K. Itagaki reported a pre-discovery detection of this transient (with an unfiltered magnitude of 17.9), obtained on 2014 January 24.85 UT with a 0.5 m reflector at the Takamizawa station, Japan. The coordinates of this transient is R.A. = $12^{\rm h}18^{\rm m}48^{\rm s}.68$ and Decl. = $+14^\circ$ 24\arcmin 43\arcsec.5 (J2000), locating at 13\arcsec.8 west and 15\arcsec.9 south of the center of the host galaxy.  The latest distance measurement of M99 reported by \citet{Tully13} as 13.9$\pm$1.5 Mpc based on the Tully-Fisher relation is adopted in this paper.  This result well matches an alternative distance reported by \citet{Poznanski09} as 14.4$\pm$2.0 Mpc via the expanding photosphere method. 

SN\,2014L was independently classified as a young SN Ic by several groups (e.g., \citealp{ Akitaya14,Li14,Ochner14,JJZhang4b}) within 1-2 days after the discovery. We thus triggered the follow-up observation campaign on the Li-Jiang 2.4 m telescope (LJT; \citealp{Fan15}) with YFOSC  (Yunnan Faint Object Spectrograph and Camera; \citealp{JJZhang14a}) at Li-Jiang Observatory of Yunnan Observatories (YNAO) and the Tsinghua-NAOC 0.8 m telescope (TNT; \citealp{wang08,TNT}) at Xing-Long Observatory NAOC. This campaign spans from $t\approx -10$ days to $t\approx +140$ days, or from $\tau \approx +3$ days to $\tau \approx +150$ days (parameter $t$ denotes to the time relative to the $V$-band maximum, $\tau$ denotes to the time compared to the shock breakout in this paper) covering the full photospheric phase. The daily observations at the first month after shock breakout makes SN\,2014L an excellent object for studying the observed properties of SNe Ic.

Figure \ref{<img>} shows the finder chart and the pre-discovery images of SN\,2014L taken by the {\it HST} Wide-Field Planetary Camera 2 (WFPC2) on Jan, 2009 under {\it HST} program GO-11966 (PI: Regan). At the distance of $D = 13.9 \pm 1.5$ Mpc, the corresponding physical pixel size in the sky is $d\sim$7 pc per pixel.  In these pre-explosion images, no point-like source can be detected within 0.\arcsec2 from the SN position. The apparent brightness of the white circle region is estimated as m$_{F336W} = 22.3\pm\,4.0$ mag, m$_{F606W} = 21.1\pm\,0.5$ mag and m$_{F814W} =20.5 \pm\,0.6$ mag.   On the other hand, there are some brighter and bluer sources located not far from the SN position (i.e., $\sim$ 0\arcsec.6).  Given the high star formation rate of M99 \citep{Soria06}, the nearby blue sources around the birth place, the larger host extinction of SN\,2014L (see the details in section \ref{subsect:Extin}), and the strong Balmer emission lines presented in the spectra (see the details in section \ref{subsect:Sp}), SN\,2014L likely comes from a relatively young stellar environment covered by  thick dust and gas.

\subsection{Photometry}
\label{subsect:Photo}

\begin{figure}
\centering
\includegraphics[width=8.5cm,angle=0]{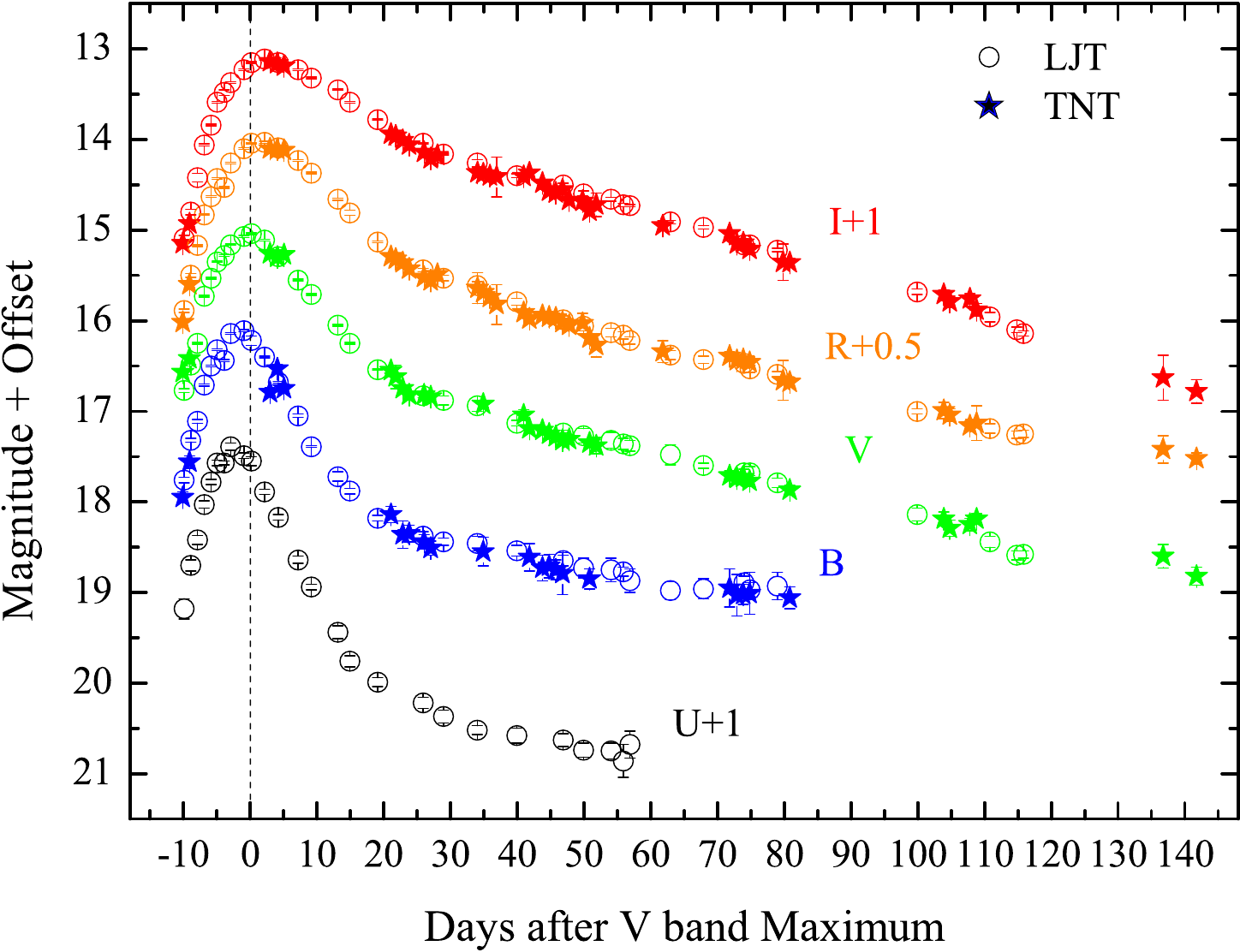}
 \caption{$UBVRI$ light curves of SN\,2014L obtained at the Li-Jiang 2.4 m telescope and $BVRI$ light curves obtained at the Tsinghua-NAOC 0.8 m telescope. An offset has been added for better visibility.}  
\label{<LC>}
\end{figure}

SN\,2014L is well-observed in the standard Johnson $UBV$ and Kron-Cousins $RI$-bands, with a typical FWHM $1\arcsec$.6 for LJT images, and $2\arcsec$.5 for the TNT images. High-quality template images were taken with the LJT+YFOSC and TNT at 2-3 years after the explosion. We perform background subtraction to remove the time-invariant background structures and better reveal the time-variant SN signals.  The photometry of this SN is transformed into the standard Johnson-Cousin photometry system (as listed in Table \ref{Tab:Pho} and presented in Figure \ref{<LC>}) through some local standard stars, as marked in the left panel of Figure \ref{<img>}. The $UBVRI$ magnitudes of these reference stars (as listed in Table \ref{Tab:Photo_stand}) are calibrated via observing a series of \citet{Landolt} photometric standard stars on three photometric nights.

%%%%%%%%%%%%%%%%%%%%%%%
\begin{deluxetable*}{lcccccccc}
\tablewidth{0pt}
%\tabletypesize{\small}
\tablecaption{$UBVRI$-band Photometry of SN\,2014L}
\tablehead{\colhead{Date}&\colhead{MJD} & \colhead{Epoch\tablenotemark{a}} & \colhead{$U$(mag)} & \colhead{$B$(mag)} & \colhead{$V$(mag)} & \colhead{$R$(mag)} & \colhead{$I$(mag)}& \colhead{Telescope} }

\startdata
Jan. 27	&	56684.72	&	-10.09 	&	\nodata	&	17.95(0.06)	&	16.57(0.02)	&	16.52(0.03)	&	16.15(0.03)	&	TNT	\\
Jan. 27	&	56684.90	&	-9.91 	&	18.18(0.11)	&	17.76(0.03)	&	16.77(0.02)	&	16.39(0.02)	&	16.09(0.03)	&	LJT	\\
Jan. 28	&	56685.74	&	-9.07 	&	\nodata	&	17.56(0.07)	&	16.42(0.02)	&	16.10(0.03)	&	15.93(0.03)	&	TNT	\\
Jan. 28	&	56685.91	&	-8.90 	&	17.70(0.06)	&	17.32(0.02)	&	16.49(0.01)	&	16.00(0.02)	&	15.80(0.03)	&	LJT	\\
Jan. 29	&	56686.92	&	-7.89 	&	17.42(0.05)	&	17.11(0.02)	&	16.25(0.01)	&	15.67(0.01)	&	15.42(0.05)	&	LJT	\\
Jan. 30	&	56687.90	&	-6.91 	&	17.03(0.04)	&	16.71(0.01)	&	15.73(0.01)	&	15.33(0.01)	&	15.06(0.01)	&	LJT	\\
Jan. 31	&	56688.95	&	-5.86 	&	16.78(0.03)	&	16.50(0.01)	&	15.53(0.01)	&	15.13(0.01)	&	14.84(0.01)	&	LJT	\\
Feb. 01	&	56689.89	&	-4.92 	&	16.57(0.03)	&	16.32(0.01)	&	15.35(0.01)	&	14.93(0.01)	&	14.59(0.01)	&	LJT	\\
Feb. 02	&	56690.94	&	-3.87 	&	16.57(0.02)	&	16.44(0.01)	&	15.28(0.01)	&	15.03(0.01)	&	14.48(0.03)	&	LJT	\\
Feb. 03	&	56691.87	&	-2.94 	&	16.39(0.04)	&	16.14(0.01)	&	15.16(0.01)	&	14.76(0.01)	&	14.37(0.01)	&	LJT	\\
Feb. 05	&	56693.90	&	-0.91 	&	16.49(0.03)	&	16.11(0.01)	&	15.07(0.01)	&	14.60(0.01)	&	14.23(0.01)	&	LJT	\\
Feb. 06	&	56694.95	&	0.14 	&	16.55(0.05)	&	16.22(0.05)	&	15.04(0.01)	&	14.54(0.01)	&	14.15(0.01)	&	LJT	\\
Feb. 08	&	56696.96	&	2.15 	&	16.89(0.03)	&	16.40(0.01)	&	15.11(0.01)	&	14.53(0.01)	&	14.11(0.01)	&	LJT	\\
Feb. 09	&	56697.79	&	2.98 	&	\nodata	&	16.79(0.05)	&	15.26(0.01)	&	14.60(0.02)	&	14.14(0.02)	&	TNT	\\
Feb. 10	&	56698.87	&	4.06 	&	\nodata	&	16.53(0.05)	&	15.30(0.01)	&	14.61(0.02)	&	14.16(0.02)	&	TNT	\\
Feb. 11	&	56699.01	&	4.20 	&	17.17(0.03)	&	16.68(0.01)	&	15.29(0.01)	&	14.59(0.01)	&	14.15(0.01)	&	LJT	\\
Feb. 11	&	56699.87	&	5.06 	&	\nodata	&	16.75(0.05)	&	15.27(0.01)	&	14.62(0.02)	&	14.19(0.02)	&	TNT	\\
Feb. 14	&	56702.00	&	7.19 	&	17.64(0.04)	&	17.05(0.02)	&	15.55(0.01)	&	14.73(0.01)	&	14.23(0.01)	&	LJT	\\
Feb. 15	&	56703.99	&	9.18 	&	17.94(0.03)	&	17.39(0.01)	&	15.71(0.01)	&	14.87(0.01)	&	14.32(0.01)	&	LJT	\\
Feb. 19	&	56707.94	&	13.13 	&	18.44(0.07)	&	17.72(0.05)	&	16.05(0.01)	&	15.16(0.01)	&	14.45(0.01)	&	LJT	\\
Feb. 21	&	56709.75	&	14.94 	&	18.76(0.06)	&	17.88(0.03)	&	16.25(0.01)	&	15.31(0.03)	&	14.59(0.01)	&	LJT	\\
Feb. 25	&	56713.89	&	19.08 	&	18.99(0.05)	&	18.18(0.03)	&	16.54(0.01)	&	15.63(0.01)	&	14.78(0.01)	&	LJT	\\
Feb. 02	&	56715.89	&	21.08 	&	\nodata	&	18.14(0.09)	&	16.55(0.02)	&	15.79(0.03)	&	14.94(0.02)	&	TNT	\\
Feb. 28	&	56716.64	&	21.83 	&	\nodata	&	\nodata	&	16.62(0.13)	&	15.82(0.05)	&	14.96(0.04)	&	TNT	\\
Mar. 01	&	56717.65	&	22.84 	&	\nodata	&	18.36(0.15)	&	16.75(0.03)	&	15.86(0.04)	&	15.01(0.03)	&	TNT	\\
Mar. 02	&	56718.62	&	23.81 	&	\nodata	&	18.35(0.09)	&	16.82(0.02)	&	15.93(0.03)	&	15.06(0.03)	&	TNT	\\
Mar. 04	&	56720.72	&	25.91 	&	19.22(0.06)	&	18.38(0.05)	&	16.83(0.02)	&	15.94(0.02)	&	15.04(0.01)	&	LJT	\\
Mar. 04	&	56720.87	&	26.06 	&	\nodata	&	18.44(0.08)	&	16.82(0.03)	&	16.02(0.03)	&	15.14(0.02)	&	TNT	\\
Mar. 05	&	56721.85	&	27.04 	&	\nodata	&	18.51(0.08)	&	16.84(0.02)	&	16.06(0.02)	&	15.21(0.02)	&	TNT	\\
Mar. 06	&	56722.85	&	28.04 	&	\nodata	&	\nodata	&	\nodata	&	15.99(0.04)	&	15.18(0.03)	&	TNT	\\
Mar. 07	&	56723.79	&	28.98 	&	19.37(0.07)	&	18.44(0.04)	&	16.88(0.05)	&	16.03(0.03)	&	15.16(0.01)	&	LJT	\\
Mar. 12	&	56728.84	&	34.03 	&	19.52(0.05)	&	18.46(0.08)	&	16.94(0.06)	&	16.11(0.04)	&	15.26(0.04)	&	LJT	\\
Mar. 12	&	56728.85	&	34.04 	&	\nodata	&	\nodata	&	\nodata	&	16.14(0.17)	&	15.36(0.04)	&	TNT	\\
Mar. 13	&	56729.71	&	34.90 	&	\nodata	&	18.55(0.16)	&	16.92(0.04)	&	16.19(0.04)	&	15.38(0.04)	&	TNT	\\
Mar. 14	&	56730.71	&	35.90 	&	\nodata	&	\nodata	&	\nodata	&	16.24(0.06)	&	15.40(0.03)	&	TNT	\\
Mar. 15	&	56731.71	&	36.90 	&	\nodata	&	\nodata	&	\nodata	&	16.32(0.22)	&	15.41(0.22)	&	TNT	\\
Mar. 18	&	56734.75	&	39.94 	&	19.58(0.08)	&	18.54(0.06)	&	17.13(0.03)	&	16.29(0.04)	&	15.40(0.02)	&	LJT	\\
Mar. 19	&	56735.76	&	40.95 	&	\nodata	&	\nodata	&	17.04(0.06)	&	16.41(0.04)	&	15.41(0.03)	&	TNT	\\
Mar. 20	&	56736.65	&	41.84 	&	\nodata	&	18.61(0.15)	&	17.19(0.03)	&	16.48(0.05)	&	15.37(0.03)	&	TNT	\\
Mar. 22	&	56738.61	&	43.80 	&	\nodata	&	18.73(0.14)	&	17.21(0.02)	&	16.44(0.04)	&	15.48(0.03)	&	TNT	\\
Mar. 23	&	56739.62	&	44.81 	&	\nodata	&	18.71(0.13)	&	17.25(0.04)	&	16.46(0.04)	&	15.57(0.03)	&	TNT	\\
Mar. 24	&	56740.62	&	45.81 	&	\nodata	&	18.75(0.11)	&	17.28(0.04)	&	16.47(0.04)	&	15.59(0.03)	&	TNT	\\
Mar. 25	&	56741.62	&	46.81 	&	\nodata	&	18.79(0.23)	&	17.32(0.04)	&	16.52(0.07)	&	15.55(0.04)	&	TNT	\\
Mar. 25	&	56741.69	&	46.88 	&	19.63(0.07)	&	18.65(0.06)	&	17.24(0.03)	&	16.49(0.02)	&	15.50(0.02)	&	LJT	\\
Mar. 26	&	56742.62	&	47.81 	&	\nodata	&	\nodata	&	17.31(0.05)	&	16.55(0.05)	&	15.67(0.03)	&	TNT	\\
Mar. 28	&	56744.62	&	49.81 	&	\nodata	&	\nodata	&	\nodata	&	16.53(0.11)	&	15.68(0.06)	&	TNT	\\
Mar. 28	&	56744.76	&	49.95 	&	19.74(0.08)	&	18.73(0.11)	&	17.27(0.08)	&	16.56(0.05)	&	15.60(0.03)	&	LJT	\\
Mar. 29	&	56745.63	&	50.82 	&	\nodata	&	18.85(0.11)	&	17.35(0.03)	&	16.69(0.04)	&	15.79(0.04)	&	TNT	\\
Mar. 30	&	56746.64	&	51.83 	&	\nodata	&	\nodata	&	17.38(0.09)	&	16.77(0.13)	&	15.72(0.13)	&	TNT	\\
Apr. 01	&	56748.84	&	54.03 	&	19.75(0.09)	&	18.75(0.13)	&	17.32(0.09)	&	16.63(0.05)	&	15.66(0.02)	&	LJT	\\
Apr. 03	&	56750.71	&	55.90 	&	19.86(0.18)	&	18.77(0.05)	&	17.36(0.06)	&	16.66(0.04)	&	15.72(0.02)	&	LJT	\\
Apr. 04	&	56751.72	&	56.91 	&	19.68(0.15)	&	18.87(0.13)	&	17.38(0.06)	&	16.72(0.04)	&	15.73(0.01)	&	LJT	\\
Apr. 09	&	56756.63	&	61.82 	&	\nodata	&	\nodata	&	\nodata	&	16.84(0.12)	&	15.95(0.05)	&	TNT	\\
Apr. 10	&	56757.78	&	62.97 	&	\nodata	&	18.98(0.10)	&	17.48(0.11)	&	16.88(0.05)	&	15.91(0.02)	&	LJT	\\
Apr. 15	&	56762.72	&	67.91 	&	\nodata	&	18.96(0.11)	&	17.60(0.03)	&	16.93(0.04)	&	15.97(0.02)	&	LJT	\\
Apr. 19	&	56766.63	&	71.82 	&	\nodata	&	18.95(0.21)	&	17.71(0.05)	&	16.89(0.05)	&	16.04(0.03)	&	TNT	\\
Apr. 20	&	56767.72	&	72.91 	&	\nodata	&	19.03(0.23)	&	17.74(0.06)	&	16.94(0.06)	&	16.15(0.05)	&	TNT	\\
Apr. 21	&	56768.69	&	73.88 	&	\nodata	&	19.02(0.11)	&	17.73(0.04)	&	16.93(0.04)	&	16.14(0.05)	&	TNT	\\
Apr. 21	&	56768.75	&	73.94 	&	\nodata	&	18.89(0.07)	&	17.68(0.04)	&	16.97(0.03)	&	16.16(0.04)	&	LJT	\\
Apr. 22	&	56769.65	&	74.84 	&	\nodata	&	19.01(0.23)	&	17.77(0.06)	&	16.96(0.06)	&	16.21(0.04)	&	TNT	\\
Apr. 22	&	56769.70	&	74.89 	&	\nodata	&	18.97(0.08)	&	17.68(0.07)	&	17.03(0.04)	&	16.16(0.01)	&	LJT	\\
Apr. 26	&	56773.77	&	78.96 	&	\nodata	&	18.93(0.15)	&	17.79(0.05)	&	17.09(0.03)	&	16.22(0.02)	&	LJT	\\
Apr. 27	&	56774.65	&	79.84 	&	\nodata	&	\nodata	&	\nodata	&	17.16(0.22)	&	16.36(0.20)	&	TNT	\\
Apr. 28	&	56775.65	&	80.84 	&	\nodata	&	19.06(0.12)	&	17.87(0.05)	&	17.18(0.05)	&	16.36(0.05)	&	TNT	\\
May 17	&	56794.72	&	99.91 	&	\nodata	&	\nodata	&	18.14(0.07)	&	17.50(0.03)	&	16.68(0.03)	&	LJT	\\
May 20	&	56798.70	&	103.89 	&	\nodata	&	\nodata	&	18.19(0.08)	&	17.49(0.09)	&	16.71(0.05)	&	TNT	\\
May 22	&	56799.61	&	104.80 	&	\nodata	&	\nodata	&	18.29(0.09)	&	17.54(0.07)	&	16.79(0.06)	&	TNT	\\
May 25	&	56802.64	&	107.83 	&	\nodata	&	\nodata	&	18.25(0.07)	&	17.66(0.05)	&	16.76(0.06)	&	TNT	\\
May 26	&	56803.64	&	108.83 	&	\nodata	&	\nodata	&	18.19(0.06)	&	17.63(0.19)	&	16.88(0.07)	&	TNT	\\
May 28	&	56805.63	&	110.82 	&	\nodata	&	\nodata	&	18.44(0.06)	&	17.69(0.05)	&	16.96(0.05)	&	LJT	\\
Jun. 01	&	56809.62	&	114.81 	&	\nodata	&	\nodata	&	18.59(0.05)	&	17.76(0.04)	&	17.10(0.03)	&	LJT	\\
Jun. 02	&	56810.65	&	115.84 	&	\nodata	&	\nodata	&	18.58(0.04)	&	17.75(0.04)	&	17.14(0.03)	&	LJT	\\
Jun. 23	&	56831.57	&	136.76 	&	\nodata	&	\nodata	&	18.60(0.13)	&	17.92(0.15)	&	17.63(0.25)	&	TNT	\\
Jun. 27	&	56836.57	&	141.76 	&	\nodata	&	\nodata	&	18.82(0.10)	&	18.02(0.06)	&	17.78(0.13)	&	TNT	
\enddata
%\tablecomments{The errors are given in brackets with units of 0.01 mag.}
\tablenotetext{a}{Relative to the $V$-band maximum, MJD = 56694.81.}
\label{Tab:Pho}

\end{deluxetable*} 
%%%%%%%%%%%%%%%%%%%%%%%

%%%%%%%%%%%%%%%%%%%%%%%
%%%%%%%%%%%%%%%%%%%%%%%

%%%%%%%%%%%%%%%%%%%%%%
\begin{deluxetable*}{cccccc}
\tablecaption{The $UBVRI$-band Magnitudes of the Local Photometric Standards in the Field of SN\,2014L}
\tablewidth{0pt}
\tablehead{\colhead{Star}&  \colhead{$U$(mag)} & \colhead{$B$(mag)} & \colhead{$V$(mag)} & \colhead{$R$(mag)} & \colhead{$I$(mag)} }
\startdata
1	&	15.05(0.02)	&	14.74(0.01)	&	14.01(0.02)	&	13.62(0.01)	&	13.24(0.02)		\\
2	&	19.61(0.04)	&	18.56(0.01)	&	17.46(0.02)	&	16.78(0.02)	&	16.17(0.03)		\\
3	&	17.29(0.03)	&	17.28(0.02)	&	16.67(0.01)	&	16.33(0.02)	&	15.96(0.02)		\\
4	&	17.51(0.03)	&	17.47(0.02)	&	16.78(0.01)	&	16.41(0.01)	&	16.02(0.02)		\\
5	&	14.30(0.02)	&	14.11(0.01)	&	13.44(0.01)	&	13.09(0.02)	&	12.73(0.02)		\\
6	&	19.03(0.03)	&	17.60(0.01)	&	16.28(0.01)	&	15.46(0.01)	&	14.72(0.01)		\\
7	&	18.36(0.03)	&	18.23(0.01)	&	17.39(0.02)	&	16.93(0.01)	&	16.44(0.02)	
\enddata
\tablecomments{See Figure \ref{<img>} for the finder chart of these reference  stars.}
\label{Tab:Photo_stand}
\end{deluxetable*}
%%%%%%%%%%%%%%%%%%%%%%%%%%%%

%%%%%%%%%%%%%%%%%%%%%%%
 \begin{figure}
\centering
\includegraphics[width=8cm,angle=0]{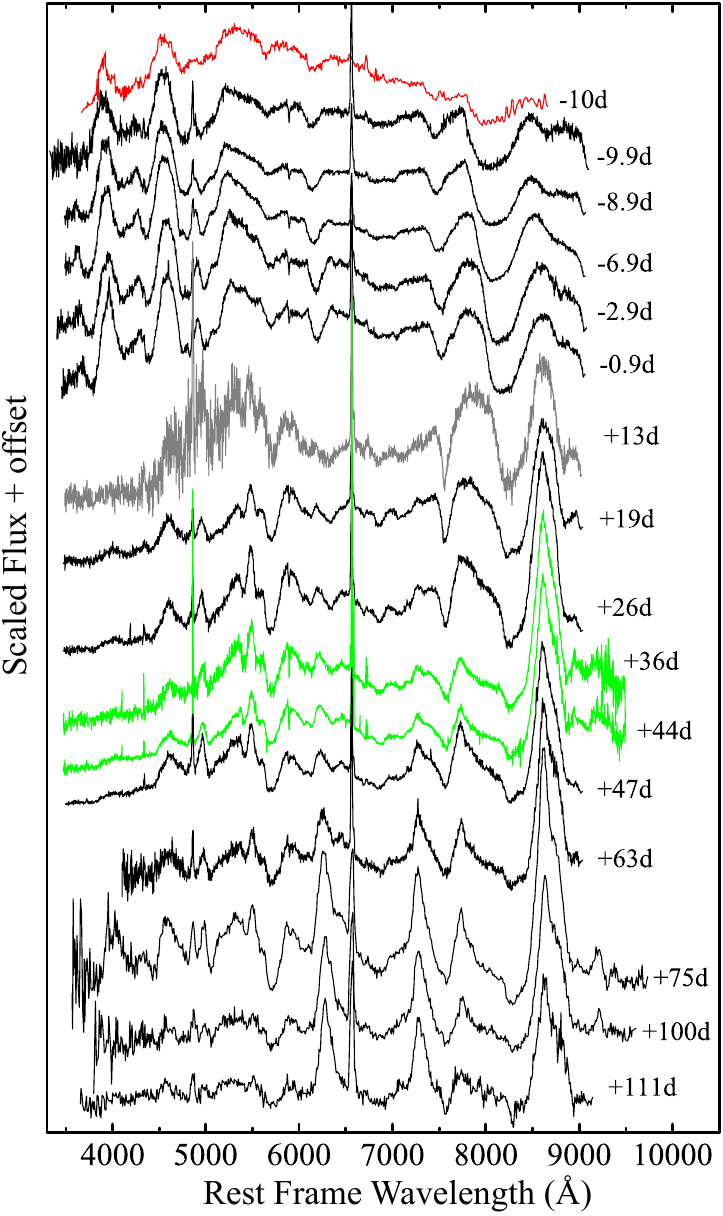}
 \caption{Spectral series of SN\,2014L.  Black, red, and green lines represent the spectra obtained by the 2.4 m Li-Jiang Telescope, the 2.16 m Xing-Long Telescope, and ANU WiFeS SuperNovA Program (AWSNAP; \citealp{Childress16}), respectively. }
\label{<Sp>}
\end{figure}
%%%%%%%%%%%%%%%%%%%%%%%

%%%%%%%%%%%%%%%%%%%%%%%
\begin{deluxetable*}{lccccccc}[!th]
%\rotate
\tabletypesize{\footnotesize}
\tablewidth{0pt}
\tablecaption{Journal of Spectroscopic Observations of SN\,2014L}

\tablehead{\colhead{Date} & \colhead{MJD} & \colhead{Epoch\tablenotemark{a}} & \colhead{Res.} & \colhead{Range} & \colhead{Exp. Time} &\colhead{Airmass}& \colhead{Telescope}\\ 
\colhead{(UT)} & \colhead{} & \colhead{(days)} & \colhead{(\AA)} & \colhead{(\AA)}  & \colhead{(s)} &\colhead{}& \colhead{(+Instrument)} } 

\startdata
Jan. 27	&	56684.81	&	-10.00 	&	3700-8700	&	16	&	3000	&	1.13	&	XLT+BFOSC	\\
Jan. 27	&	56684.91	&	-9.90 	&	3380-9150	&	18	&	2100	&	1.04	&	LJT+YFOSC	\\
Jan. 28	&	56685.91	&	-8.90 	&	3520-9130	&	18	&	2400	&	1.04	&	LJT+YFOSC	\\
Jan. 30	&	56687.91	&	-6.90 	&	3520-9130	&	18	&	2100	&	1.04	&	LJT+YFOSC	\\
Feb. 03	&	56691.88	&	-2.93 	&	3430-9150	&	18	&	1800	&	1.03	&	LJT+YFOSC	\\
Feb. 05	&	56693.90	&	-0.91 	&	3480-9130	&	18	&	2400	&	1.06	&	LJT+YFOSC	\\
Feb. 19	&	56707.90	&	13.09 	&	3520-9100	&	18	&	2400	&	1.15	&	LJT+YFOSC	\\
Feb. 25	&	56713.90	&	19.09 	&	3500-9100	&	18	&	2100	&	1.21	&	LJT+YFOSC	\\
Mar. 04	&	56720.72	&	25.91 	&	3500-9100	&	18	&	2500	&	1.09	&	LJT+YFOSC	\\
Mar. 25	&	56741.70	&	46.89 	&	3520-9100	&	18	&	2700	&	1.04	&	LJT+YFOSC	\\
Apr. 10	&	56757.79	&	62.98 	&	3530-9150	&	18	&	3000	&	1.25	&	LJT+YFOSC	\\
Apr. 22	&	56769.76	&	74.95 	&	3590-9800	&	50	&	1500	&	1.29	&	LJT+YFOSC	\\
May 17	&	56794.70	&	99.89 	&	3920-9680	&	50	&	1500	&	1.32	&	LJT+YFOSC	\\
May 28	&	56805.65	&	110.84 	&	3690-9210	&	50	&	1800	&	1.21	&	LJT+YFOSC	
\enddata
\tablecomments{Spectroscopic observations of SN\,2014L.}
\tablenotetext{a}{Relative to the $V$-band maximum, MJD = 56694.81.}
\label{Tab:Spec_log}
\end{deluxetable*}

%%%%%%%%%%%%%%%
\begin{deluxetable}{ccccc}[!th]
\tablewidth{0pt}
\tablecaption{Absolute peak brightness of SN\,2014L in $UBVRI$-bands}
\tablehead{\colhead{Band}  & \colhead{$t_{\rm max}$} & \colhead{$m_{\rm peak}$}& \colhead{$M_{\rm peak}$}  & \colhead{$\Delta m_{15}$}\\
\colhead{} & \colhead{(MJD-56000)} & \colhead{(mag)} &  \colhead{(mag)}&  \colhead{(mag)}  }
\startdata
$U$   & 692.45(0.40) & 16.45(0.04) &-17.54(0.41)&2.00(0.05) \\
$B$  & 693.16(0.25) & 16.16(0.03)  &-17.30(0.35)&1.60(0.03)\\
$V$   & 694.81(0.20) & 15.06(0.02) &-17.73(0.28)&1.10(0.03)\\
$R$   &696.88(0.20) & 14.54(0.02) &-17.71(0.23)&0.94(0.02)\\
$I$   & 697.33(0.25) & 14.14(0.03) &-17.64(0.19)& 0.56(0.03)
\enddata
\label{Tab:LV_par}
\end{deluxetable}

\subsection{Spectroscopy}
\label{subsect:Sp}
Figure \ref{<Sp>} shows the spectral sequence of SN\,2014L over a period of about four months starting from 2014 January 27. The observation journal of these spectra is listed in Table \ref{Tab:Spec_log}, including 13 spectra from the LJT (+YFOSC) and 1 spectrum from the Xing-Long 2.16 m telescope (XLT) with the BFOSC (Beijing Faint Object Spectrograph and Camera). All of these spectra are calibrated in both wavelength and flux, and  they are corrected for telluric absorption and redshift. The narrow emission lines (e.g., H$\alpha$, H$\beta$) presented in the spectra are due to the contamination in the host galaxy, which becomes dominating at late phase. Variations of these narrow emissions are due to the slit position and spectral resolution. Two spectra from the ANU WiFeS SuperNovA Program (AWSNAP; \citealp{Childress16}) are also included in this figure to fill the observation gaps.

\begin{figure}
\centering
\includegraphics[width=8.5cm,angle=0]{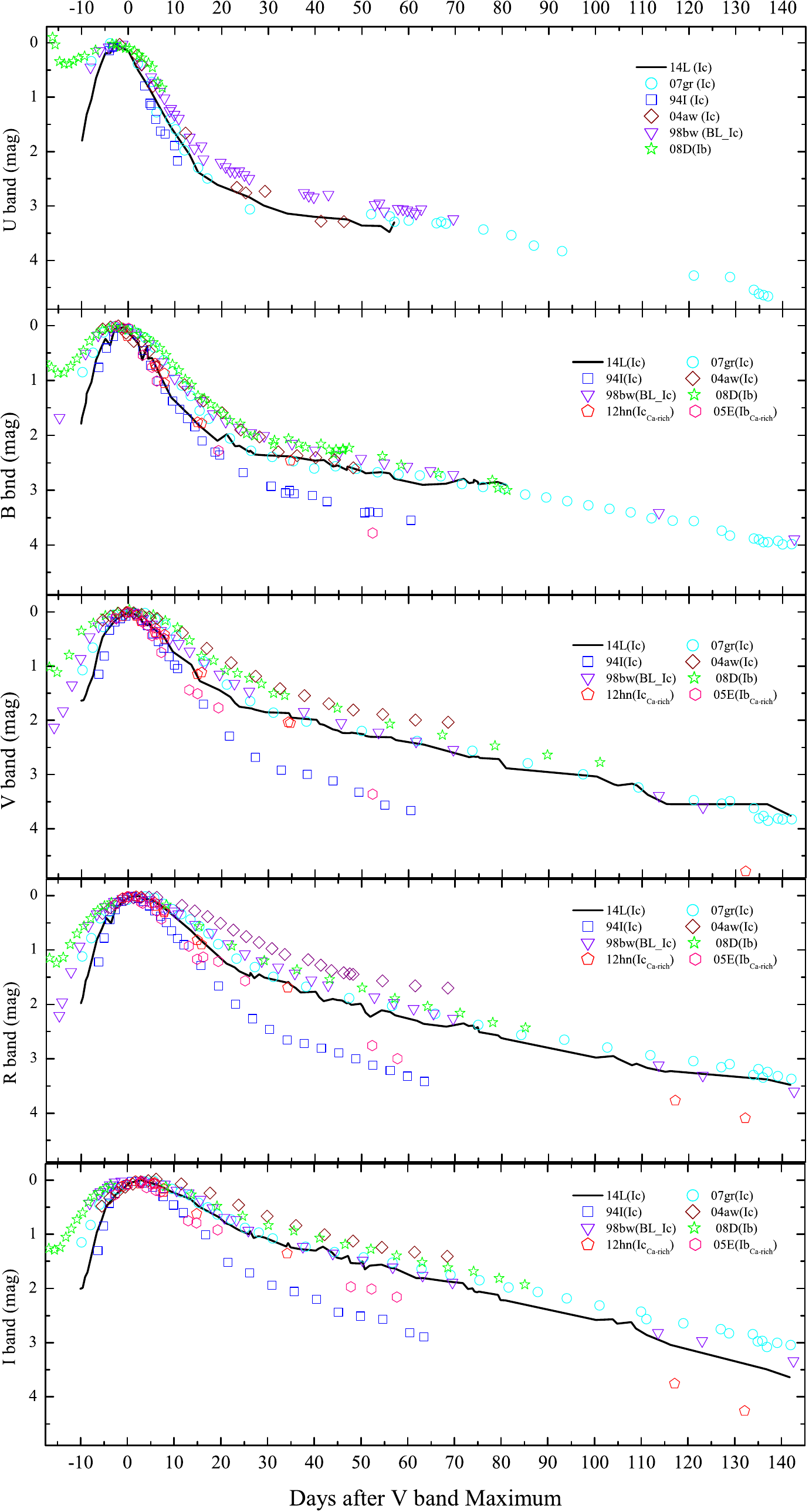}
 \caption{Comparison of the $UBVRI$ light curves of SN\,2014L with some well-observed SNe Ib/c, including SN\,2007gr, SN\,1994I, SN\,2004aw, SN\,1998bw, SN\,2008D, SN\,2012hn, and SN\,2005E. All light curves are normalized to their peaks in each band. }
\label{<LC_comp>}
\end{figure}
%%%%%%%%%%%%%%%%%%%%%%%

\section{Photometric Results}
\label{sect:LC}
%%%%%%%%%%%%%%%%%%%%%%%

We fit the UBVRI light curves with polynomials to derive the peak magnitudes, dates, and the post-peak decline rates, as listed in Table \ref{Tab:LV_par}. For example, SN\,2014L reached its $V-$band peak on MJD = 56694.81$\pm1.0$, and it reached the peaks at slightly different epochs in other bands. By adopting the distance $D=13.9\pm1.5$ Mpc (as discussed in section \ref{sect:Obs}), and the total extinction $E(B-V)= 0.67 \pm\ 0.11$ (derived in section \ref{subsect:Extin}), we obtain an absolute $V$-band peak magnitude of $-17.73\pm0.28$ mag for SN\,2014L.

\subsection{Light Curves}
\label{subsect:PP}

%%%%%%%%%%%%%%%%%%%%%%%
Figure \ref{<LC_comp>}  displays the $UBVRI$ light curves of SN\,2014L compared to those of  some typical SNe to better understand the photometric  properties of SN\,2014L. Three SNe Ic, SN\,1994I \citep{Richmond96}, SN\,2007gr \citep{Valenti08,Chen14}, and SN\,2004aw \citep{Taubenberger06};  the board-line SN Ic, SN\,1998bw \citep{Clocchiatti11}; SN Ib, SN\,2008D \citep{Modjaz09,Tanaka09,Bianco14}; faint and calcium-rich events SN\,2005E \citep{Perets10} and SN\,2012hn\citep{Valenti14} are plotted in Figure \ref{<LC_comp>}. 

The light curves of SN\,2014L are overall similar to those of SN\,2007gr, though the former appears slightly narrower than the latter. It is surprising that SN\,2014L, SN\,2004aw, SN\,2007gr, and SN\,1998bw show similar evolution in the $U$-band, with a small scatter of $<$0.2 mag. In the $B$-band, SN\,2014L, SN\,1994I and SN\,2012hn also show similar light-curve evolution at $t < + 20$ days. After that, these SNe exhibit different decay rates. Larger scatters also exist in the $VRI$-bands at similar phases.

These SNe can be divided into three clusters, depending on the decline rate from $0<t<+30$ days.  For example, the slow declining group including SN\,2004aw, SN BL-Ic 1998bw, SN\,Ib 2008D; the fast declining group including SN Ic 1994I and SN Ca-rich 2005E; and the intermediate declining group including SN Ic 2014L, SN Ic 2007gr, and SN Ca-rich 2012hn.  In each cluster, however, conspicuous discrepancies can be identified in spectral comparisons at both photospheric and nebular phase of Section \ref{sect:SP}. 

The relatively similar peak properties are often in contrast with the considerable spread in the properties of the late-time tail of the light curve (e.g., \citealp{Wheeler15}, hereafter is \B{WJC 15}). Clear differences existing among the tails of these light curves might be related to the different optical opacities and  $\gamma$-ray leakage rates of different SNe Ib/c.

%%%%%%%%%%%%%%%%%%%%%%%
 \begin{figure*}
\centering
\includegraphics[width=17cm,angle=0]{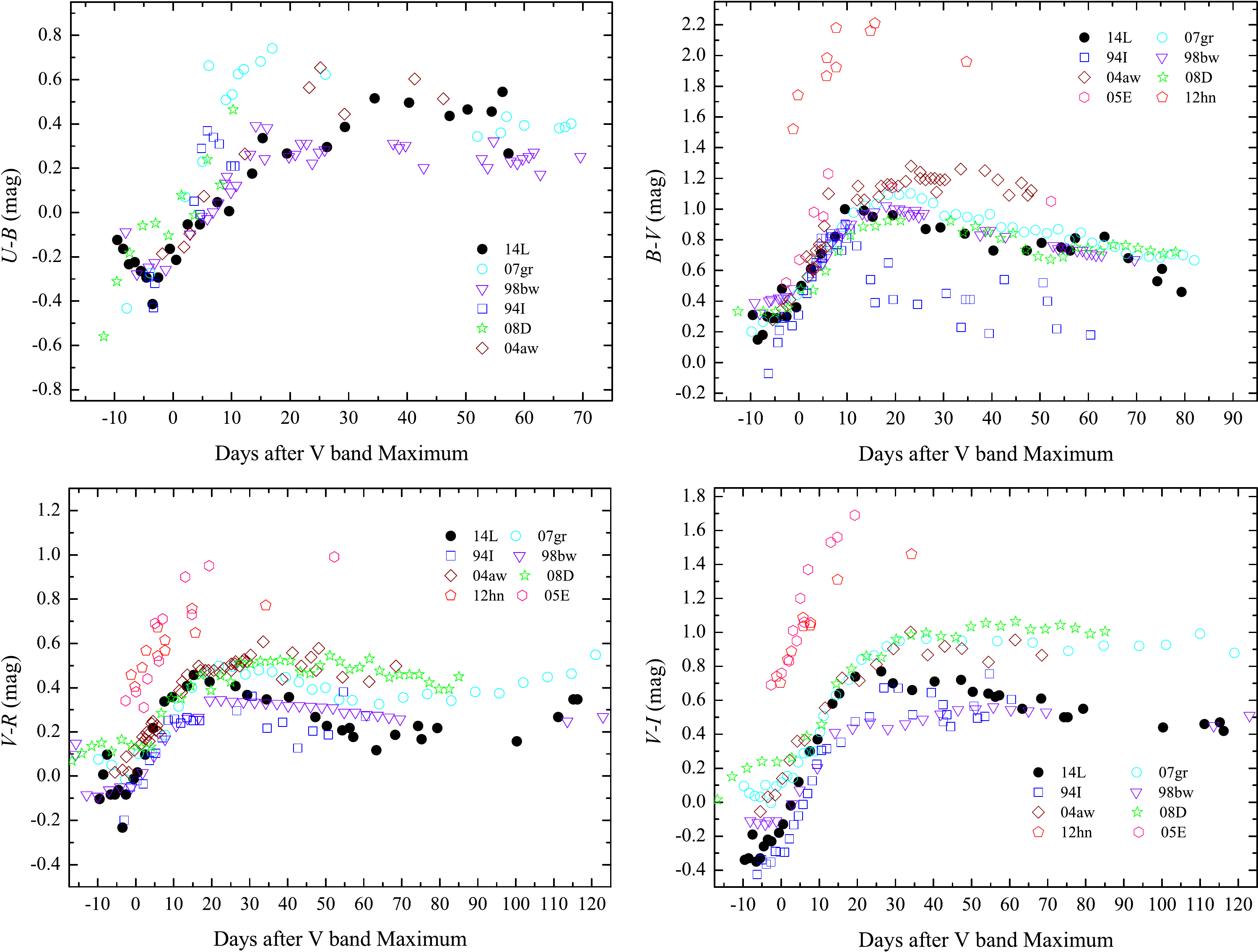}
 \caption{Comparison of the color curves among SN\,2014L, SN\,2007gr, SN\,2012hn, SN\,1998bw, SN\,2004aw, SN\,1994I, and SN\,2008D. All of these curves are corrected for the reddening due to the Milky Way and Host galaxies.}
\label{<CC>}
\end{figure*}
%%%%%%%%%%%%%%%%%%%%%%%

\subsection{Reddening}
\label{subsect:Extin}

The early spectra of SN\,2014L show red continuum and significant absorption of narrow \NaI\ from the host galaxy.  These features suggest substantial line-of-sight reddening towards the SN.  The equivalent width (EW) of \NaI\,  absorption in the SN spectra can be used to estimate the reddening according to some empirical correlations between reddening and $EW$ of \NaI, e.g., $E(B-V) = 0.16 EW_{\rm Na} - 0.01$ \citep{Tura03}, and $E(B-V) = 0.25 EW_{\rm Na}$ \citep{Barbon90}. However, these empirical correlations usually exhibit large scatter for the extinction measurement (e.g., \citealp{Poznanski11,Phillip13}). In SN\,2014L, EW(Na I D) is measured as 2.7 $\pm 0.1$ \AA\, which corresponds to a color excess of $E(B-V)_{\rm host}$ = 0.42$\pm$0.04 or 0.68$\pm$0.05 following the above relations.

Alternatively, the host-galaxy reddening of SNe Ib/c can be estimated by a photometric method. Based on a larger sample of SNe Ib/c, \citet{Drout11} found that the $V-R$ color of extinction-corrected SNe Ib/c is tightly clustered at 0.26$\pm$0.06 mag at t$\approx+10$ days after the $V$-band maximum, and $0.29\pm0.08$ mag at $t \approx +10$ days after the $R$-band maximum. This method yields an estimate of $E(B-V)_{\rm total}=0.75\pm0.05$ mag for SN\,2014L  assuming an $R_V = 3.1$ Milky Way extinction law for the host galaxy. Adopting the Galactic extinction $E(B-V)_{\rm Gal} = 0.04\pm0.01$ \citep{Schlafly11}, the host reddening derived from the $V-R$ color is $E(B-V)_{\rm host} = 0.71 \pm0.05$ mag.

Combining the estimations obtained from \NaI\ absorption and $V-R$ color, an average value of the total reddening, $E(B-V)=0.67\pm0.11$ mag, is adopted in this paper.

\subsection{Color curves}
\label{subsect:CC}

The reddening-corrected color curves of SN\,2014L are displayed in Figure \ref{<CC>}, together with those of the comparison sample shown in Figure \ref{<LC_comp>}. At early time, the $B -V$ color evolution of SN\,2014L is similar to that of SN\,1998bw, SN\,2008D, and SN\,2007gr, however, deviations can be identified in other bands especially at t $> +$20 days relative to the V-band maximum. The scatter shown before the peak can be attributed to different temperature and metal abundance in the outer ejecta, while the scatter seen after $t\approx+20$ days could be related to the duration of photospheric phase.

The large scatter among the colors of these SNe, even in the SNe Ib/c group, might not suggest a uniform color evolution for these SNe. However, as mentioned before, there seems to exist an intrinsic $V-R$ color for regular SNe Ib/c at about ten days after the peak \citep{Drout11}. Such an inherent color distribution might also exist in $B-V$ and $V-I$ color at a similar phase in Figure \ref{<LC_comp>}

%%%%%%%%%%%%%%%%%%%%%%%
\begin{figure}
\centering
\includegraphics[width=8.5cm,angle=0]{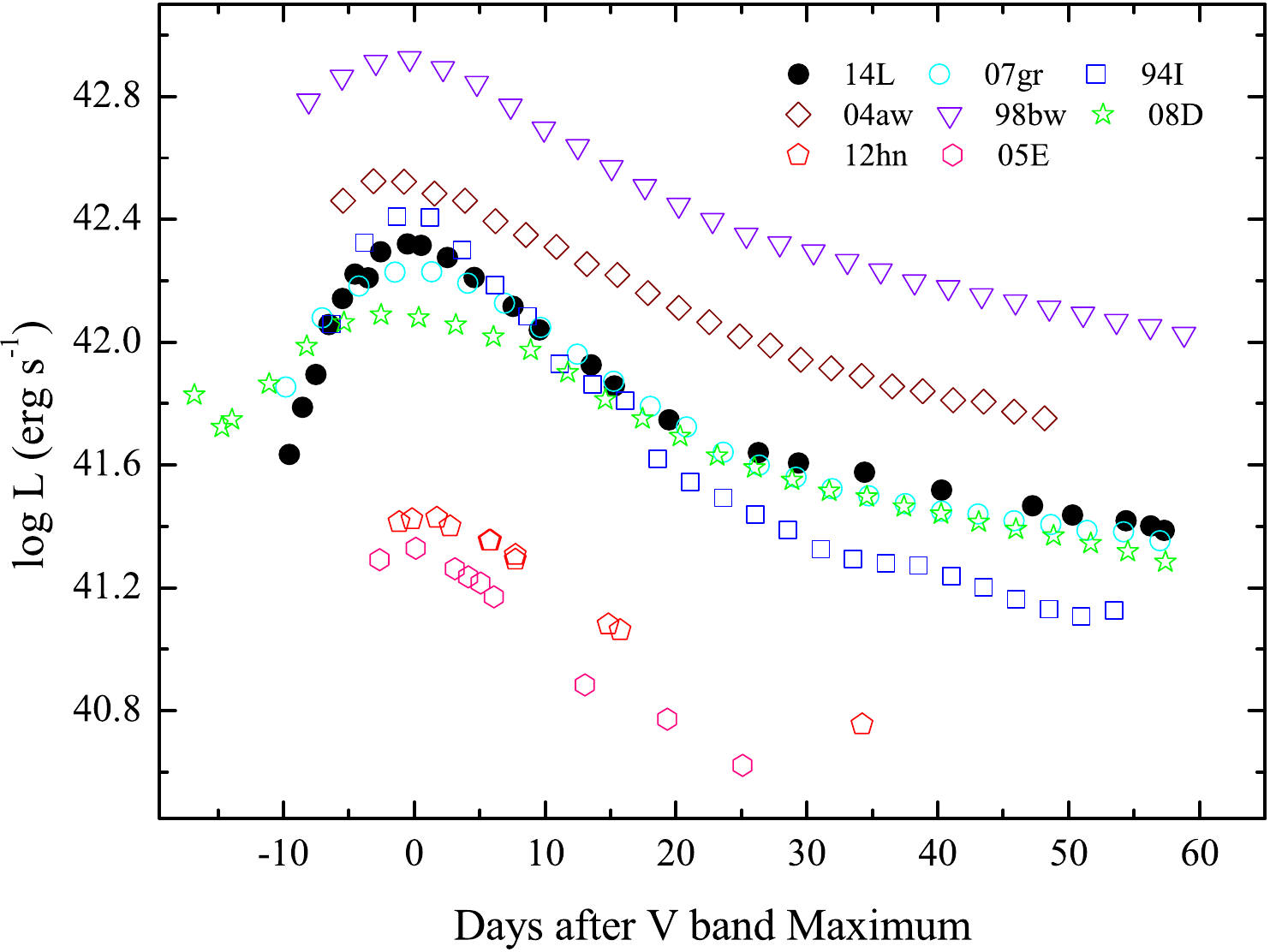}
 \caption{$UBVRI$ quasi-bolometric light curves of SN\,2014L, SN\,2007gr, SN\,2005E, SN\,2012hn, SN\,1998bw, SN\,2004aw, SN\,1994I, and SN\,2008D. }
\label{<bolo>}
\end{figure}
%%%%%%%%%%%%%%%%%%%%%%%

\subsection{Quasi-bolometric light curve}
\label{subsect:Bolo}

Figure \ref{<bolo>} displays the quasi-bolometric light curves of our sample, based on the $UBVRI$ photometry presented in Figure \ref{<LC_comp>}. SN\,2014L researches the peak quasi-bolometric luminosity of $L = (2.06 \pm\ 0.50) \times 10^{42}$ erg s$^{-1}$ at MJD = 56693.86 which is about one day earlier than the $V$-band maximum. The fractional contribution of the ultraviolet (UV) and near-infrared (NIR)  emission to the bolometric luminosity can be $\sim30-40\%$ at around the peak for SNe Ib/c (e.g., $\sim40\%$ in SN\,2008D; \citealp{Modjaz09}). Considering the missed UV and NIR flux of SN\,2014L, the total bolometric flux of this SN could reach to $\sim$ 3$\times10^{42}$ erg s$^{-1}$ around the peak.

\section{Spectral Evolution}
\label{sect:SP}

\begin{figure*}
\centering
\includegraphics[width=17cm,angle=0]{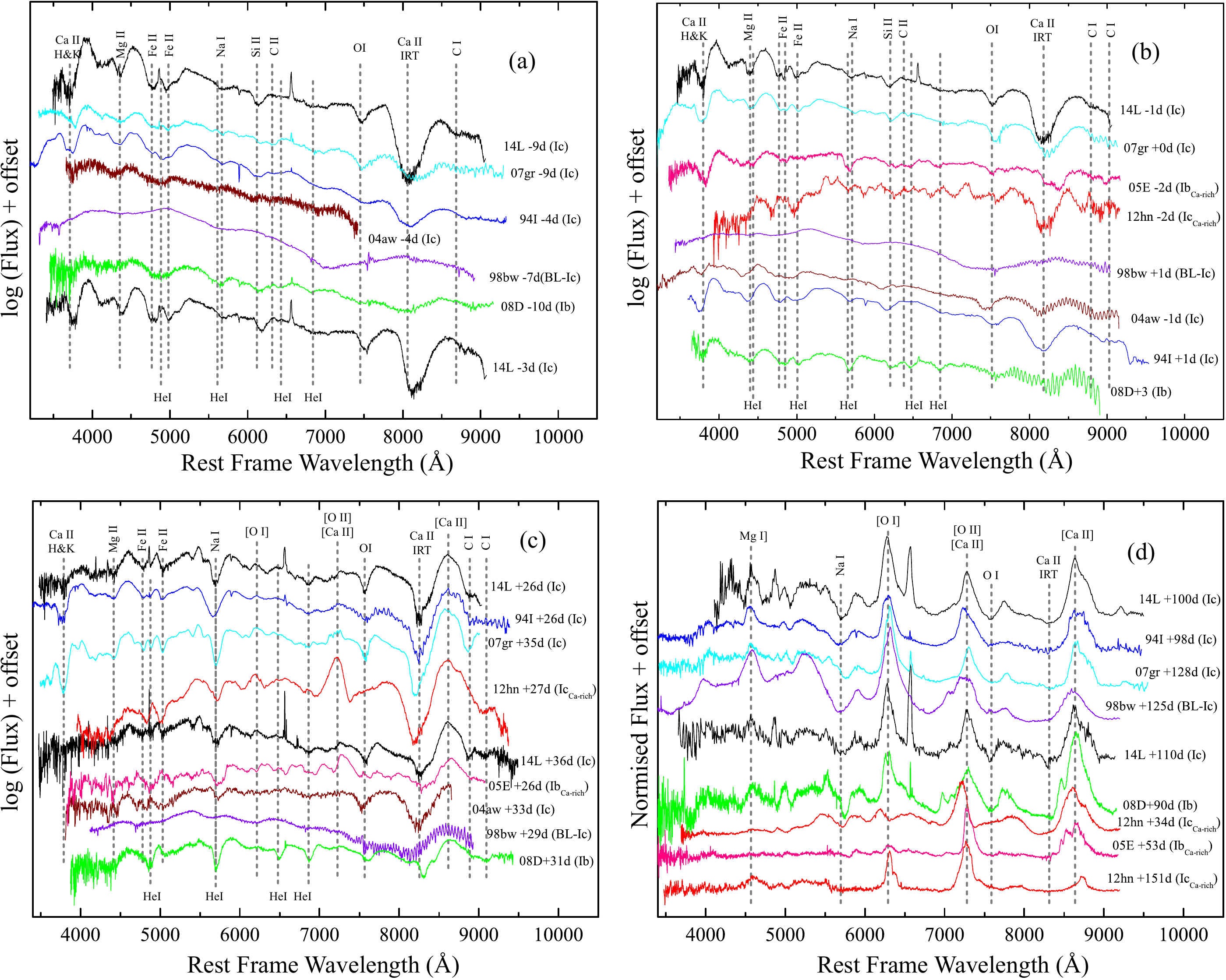}
 \caption{Spectral comparison among SN\,2014L, SN\,2007gr, SN\,2005E, SN\,2012hn, SN\,1998bw, SN\,2004aw, SN\,1994I and SN\,2008D at four selected phases.  The positions of major line features are marked by dashed-lines referring to the spectral identification of SN\,2007gr \citep{Valenti08,Hunter09,Chen14}. The spectra in panel (d) are normalized to the flux at around 7300\AA\ and vertically shifted  for better visibility. }
\label{<Spcomp>}
\end{figure*}

The spectral sequence of SN\,2014L, covering the full evolution at the photospheric and a part of the early nebular phase, was {presented} in Figure \ref{<Sp>}. A closer comparison between SN\,2014L and other representative SNe Ib/c are presented at four selected phases as plotted in Figure \ref{<Spcomp>}.  All of these spectra have been corrected for redshift and reddening.

\subsection{Early phase}
\label{subsect:spear}

The early-time spectra (i.e., at a few days after the explosion) contain prominent features to distinguish it among different subclasses of SNe. Figure \ref{<Spcomp>}a shows the t$\sim$ $-$7 day spectra of SN\,2014L and some comparison SNe Ib/c including SN\,2007gr \citep{Chen14}, SN\,1994I \citep{Filippenko95}, SN\,2004aw \citep{Modjaz14}, SN\,1998bw \citep{Patat01}, and SN\,2008D \citep{Modjaz09}. Compared to the almost featureless spectrum of SN\,2004aw and the board-lined spectrum of SN\,1998bw, the spectra of SN\,2014L, SN\,2007gr, and SN\,1994I are characterized by the absorptions of some intermediate-mass elements (IMEs, e.g., Ca, Si, O, and C) and \FeII. The clear absorption near 6150 \AA\ in the spectra of SN\,2014L, SN\,2007gr, and SN\,1994I can be identified as the feature of \SiII\ \ld6355.  The absorption near about 5600\AA\ is likely attributed to the absorption of \NaI\ rather than \HeI\ \ld5876 because of the absence of other helium lines as found in the spectrum of SN Ib 2008D. Overall, SN\,2014L shows close resemblances to SN\,2007gr and SN\,1994I, except that it has more prominent absorption features of \CaII\ NIR triplet. The difference in absorptions of \CaII\ NIR triplet might indicate difference at the outer layer of ejecta, as discussed in section \ref{subsect:vel}.

\subsection{Around Maximum}
\label{subsect:spmax}

\begin{figure}
\centering
\includegraphics[width=8.5cm,angle=0]{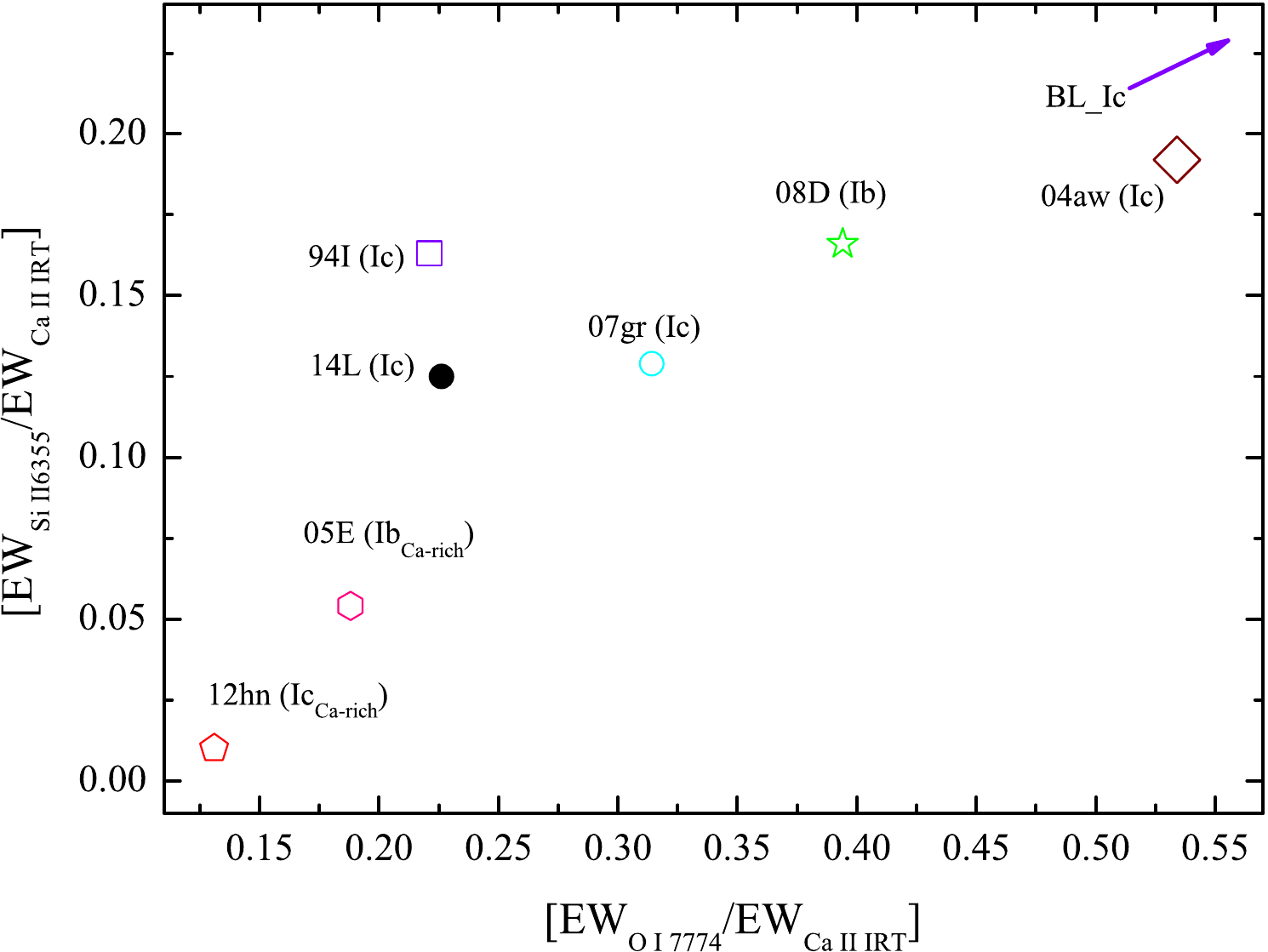}
 \caption{The EW ratio of  \OI\ \ld 7774 and \CaII\, NIR triplet is plotted versus that of \SiII\ \ld6355 and \CaII\, NIR triplet for sample shown in Figure \ref{<Spcomp>}. The arrow symbol represents the region expected for SN\,1998bw. The size of each symbol are related to the estimated uncertainties.  }
\label{<SiOCa>}
\end{figure}

At around the maximum, the spectral features of SN\,2014L are very similar to those of SN\,2007gr. These two SNe Ic show narrower absorptions of Si, O, Na, and Fe than SN\,1994I and SN\,2004aw. In particular, SN\,2014L seems to have unusually strong absorption features for a \CaII\ NIR triplet, comparable to that of the Ca-rich events such as SN\,2005E \citep{Perets10} and SN\,2012hn \citep{Valenti14}. The strong \CaII\ NIR triplet in SN\,2014L might relate to the abundance enrichment of IMEs at the outer layer. To investigate the relative abundance of Si, Ca, and O measured from the spectra shown in Figure \ref{<Spcomp>}b, we plot the EW ratio of \OI\ \ld7774 / \CaII\ NIR triplet versus \SiII\ \ld6355 / \CaII\ NIR triplet in Figure \ref{<SiOCa>}. One can see that the absorption lines ratio Ca/O of SN\,2014L is close to that of SN\,1994I, while the Ca/Si ratio of SN\,2014L is similar to that of SN\,2007gr. Both of these two ratios measured from SN\,2014L are smaller than the Ca-rich events like SN\,2005E and SN\,2012hn, ruling out the Ca-rich subclass. Nevertheless, the synthetic spectra at around maximum by \citep{Iwamoto98} suggests that the strength of \CaII\ NIR triplet is weaker than that of \SiII\ and \OI.

Compared to other subclasses of SNe Ib/c, the BL SNe Ic tend to cluster at the top right region of Figure \ref{<SiOCa>}. Note, however,  it is hard to measure the absorptions of \CaII\ NIR triplet and \OI\ \ld7774 in the spectra of SN\,1998bw because these two features are blended before and near the maximum light. Moreover, these absorptions cover the wavelength range of the high-velocity \CaII\ NIR triplet and the photospheric components of \CaII\ NIR triplet and \OI\ \ld7774.

According to the distribution show in Figure \ref{<SiOCa>} and Figure \ref{<bolo>}, we can divide our sample of SNe Ib/c into three groups.  For example, SN\,2005E and SN\,2012hn had lower luminosity and located at the bottom of Figure  \ref{<SiOCa>}.   The SN\,1998bw-like objects tend to have a smaller amount of calcium and higher luminosity, while the rest of the SNe Ib/c in our sample have a moderate luminosity and lie in the center of Figure  \ref{<SiOCa>}.  In our sample of SNe Ic, SN\,2004aw can be regarded as the transitional event linking the BL SNe Ic with the ordinary ones. The distribution of SNe Ic in Figure \ref{<SiOCa>} and Figure \ref{<bolo>} suggests that the relative strength of Ca II NIR triplet measured in the near-maximum spectra is inversely related to their peak luminosities. Namely, SNe Ic with a significant amount of calcium tend to have lower luminosities. This result may still suffer from small-number statistics, but it is consistent with the theoretical prediction that the O/Ca ratio increases with the mass of progenitor for core-collapse SNe \citep{HouckFransson}.
Thus, a comparable O/Ca ratio seen in SN\,2014L, SN\,1994I, and SN\,2007gr might imply that they have similar progenitor masses and brightness, while SN\,2004aw and SN\,1998bw have likely arisen from more massive progenitors. It is possible that the Ca-rich events might have lower-mass progenitor systems given the fact that they tend to occur at positions far from the host galaxies.  However, massive stars could go for an outing through tidal stripping during galaxy interactions.  And we can not rule out a massive progenitor  scenario for the Ca-rich events thoroughly \citep{Kasliwal12}.

%%%%%%%%%%%%%%%%%%%%%%%

\subsection{One month after maximum}
\label{subsect:sponem}

By t $\sim$ 1 month, the spectra show a clear distinction for different SNe Ic, as shown in Figure \ref{<Spcomp>}c. For example, the t$\approx$+26 day spectrum of SN\,2014L shows a close resemblance to the $t \approx+35$ day of SN\,2007gr and the $t \approx+26$ day spectrum of SN\,1994I, while its t$\approx$+36 day spectrum shows less photospheric features compared to the $t \approx+35$ day spectrum of SN\,2007gr. This implies that SN\,2014L has a relatively faster spectral evolution than SN\,2007gr. In comparison, the Ca-rich class seems to be the most rapidly evolving spectral feature. At this phase, some nebular features, such as the emission at around 7300\AA\ (due to the blended lines of \OII\ and \CaII) started to emerge. The significant differences in the spectral evolution conform to the scatter seen in the color curves at similar phases. The absence of obvious \OII\ \ld7774 absorption in the spectra of Ca-rich SN\,2012hn and SN\,2005E might relate to the small ratio of O/Ca of these events.

It should be noted that the \CI\ \ld9087 absorption is detected in the spectra of SN\,2014L and SN\,2007gr. This carbon feature becomes unambiguous in the $t\sim$26 day observed spectrum of SN\,1994I due to the diffraction fringing effect of the instrument. However, detection of this \CI\ line was reported in the $t\approx+10$ day spectrum of SN\,1994I \citep{Baron96}. The feature of \CI\ observations indicates that SN\,2014L, SN\,2007gr, and SN\,1994I might arise from carbon-rich progenitors \citep{Baron96,Valenti08}.

\subsection{Beginning of Nebular Phase}
\label{subsect:spneb}

Figure \ref{<Spcomp>}d displays the spectra of these SNe at the beginning of nebular phase when the spectra of core-collapse SNe are dominated by a few emissions of IMEs (e.g., O, Mg, Ca). For example, the latest spectrum of SN\,2014L is dominated primarily by the emission lines [\OI] \ld\ld6300, 6364, a blend of [\CaII] \ld\ld7291, 7324 and [\OII] \ld\ld7300, 7330, and the \CaII\ NIR triplet. However, some photospheric components (e.g., the absorptions of \NaI, \OI\ \ld7774 and \CaII\ NIR triplet) are still evident in this phase. Similar features are also found in the spectra of SN\,1994I, and SN\,2007gr. The BL-Ic SN\,1998bw shows similar emissions but contains more photospheric features,  indicating a slower evolution in the broad-lined event.  

The spectrum of SN\,2012hn at $t\approx+34$ days keeps some photospheric features and  dominated by the emissions of [\CaII] \ld\ld7291, 7324 and [\OII] \ld\ld7300, 7330, and the \CaII\ NIR triplet.  Similar emissions are also found in the spectrum of SN\,2005E at $t\approx+53$ days with less photospheric components.   The difference between  Ca-rich event, and SNe Ib/c at the early nebular era is the emission of [\OI] \ld\ld6300, 6364 which is very weak in the former. Besides, SN\,2005E did not develop clear emission of Mg I] and SN\,2012hn develops this feature rather slowly. These imply different progenitor systems or explosion mechanism between SNe Ib/c and Ca-rich events.  
The spectrum of SN\,2012hn at $t\approx$+151 days evolved to the later nebular phase is similar to the spectra of SNe Ic at about one year after the explosion. It confirms the fact of fast evolution in Ca-rich events.

The apparent and almost symmetric double-peaked feature is found in the  [\OI] \ld\ld6300, 6364 emissions of SN\,2008D. The trace of double-peaked [\OI]  might exist in that of SN\,2007gr and SN\,1998bw. However, it is not clear in the spectra of SN\,2014L. These might indicate a different geometric structure of the ejecta.

%%%%%%%%%%%%%%%%%%%%%%%
\begin{figure}
\centering
\includegraphics[width=8.5cm,angle=0]{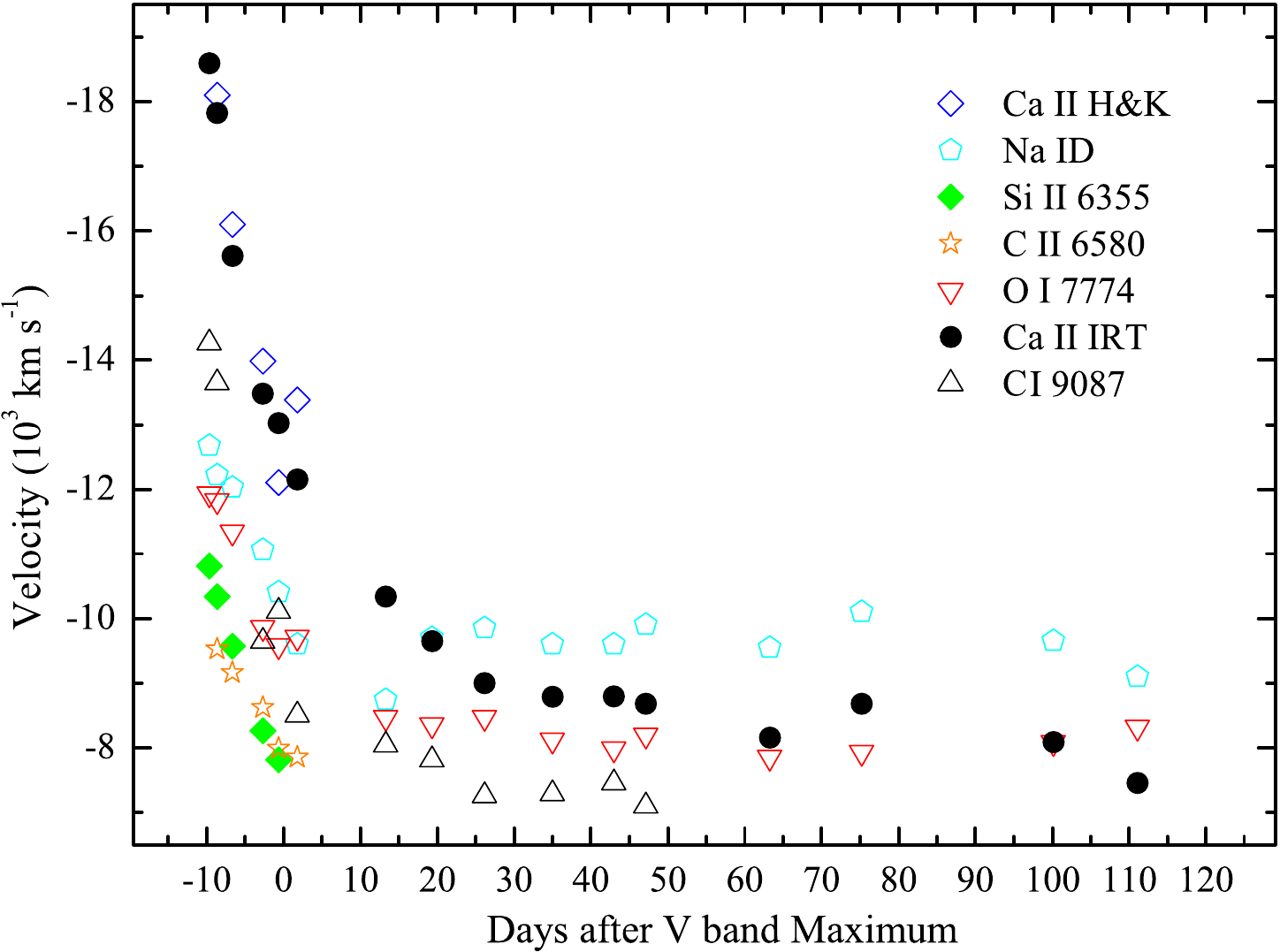}
 \caption{Ejecta velocities of SN\,2014L, as inferred from absorption minima of \CaII\ H\&K, \NaI, \SiII\ \ld6355, \CII\ \ld6580, \OI\ \ld7774, \CaII\ NIR triplet, and \CI\ \ld9087. }
\label{<Vel>}
\end{figure}

\begin{figure}
\centering
\includegraphics[width=8.5cm,angle=0]{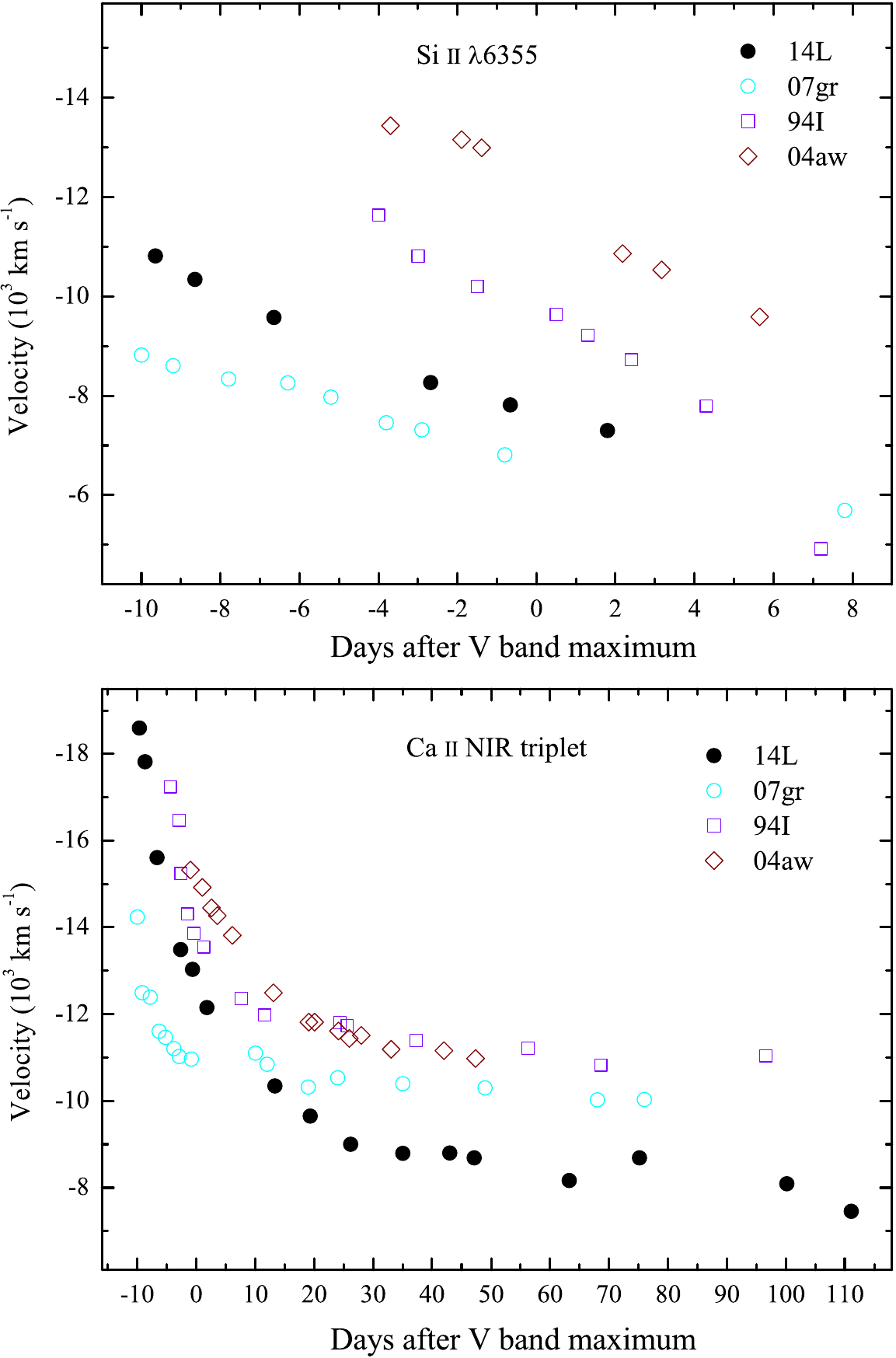}
 \caption{Velocity comparison of SN\,2014L, SN\,2007gr, SN\,1994I, and SN\,2004aw. Top panel: velocity inferred from  absorption minimum of \SiII\ \ld6355; Bottom panel: average velocity inferred from absorption minimum of  \CaII\ NIR triplet. }
\label{<Vel_Comp>}
\end{figure}
%%%%%%%%%%%%%%%%%%%%%%%

\subsection{Velocities}
\label{subsect:vel}

The ejecta velocities derived by the absorption minima of \CaII\ H\&K, \NaI, \SiII\ \ld6355, \CII\ \ld6580, \OI\ \ld7774, \CaII\ NIR triplet, and \CI\ \ld9087 of SN\,2014L are shown in Figure \ref{<Vel>}. In the early phase, the \CaII\ shows a much higher velocity relative to Na~{\sc i}, \SiII, \CII,  and \OI\ , and this discrepancy is similarly seen in type Ia supernovae (i.e., \citealp{zhao15}). This might be due to the fact that Ca II has relatively lower excitation/ionization energy and it can be formed at more distant regions (higher velocities). At $t>+10$ days, the velocities of \CaII, Na~{\sc i}, and \OI\ decline slowly and they all tend to reach a velocity plateau of about $8000-10,000$\kms. We note that the velocity of \NaI\ shows a significant increase from 8500\,\kms\ to 10000\,\kms\ during the period from t$\sim$10 days to t$\sim$20 days. This is seen in some SNe Ic (e.g., \citealp{Hunter09}) and might be due to the contributions from other spectral lines (e.g., \NaII\ or \HeI\ \ld5876). 

Figure \ref{<Vel_Comp>} shows the comparison of the velocities inferred from\SiII\ \ld6355 and  \CaII\ NIR triplet for different SNe Ic, including SN\,2014L, SN\,2007gr, SN\,1994I, and SN\,2004aw. One can see that SN\,2014L shows a higher \SiII\ velocity than that of SN\,2007gr. Before the maximum light, the velocity of the \CaII\ lines in SN\,2014L exhibits a remarkably high decline rate $\sim$ 600 km s$^{-1}$ day $^{-1}$.  The  more significant decline rate of \CaII\ velocity makes a lower velocity plateau of SN\,2014L than the comparison.

%%%%%%%%%%%%%%%%%
\begin{deluxetable}{lccccc}

\tablecaption{Explosion Parameters of Some Well-observed SNe Ic}
\tablehead{\colhead{SN} & \colhead{$t_r$\tablenotemark{a}} & \colhead{$L_{\rm peak}$} & \colhead{$v_{\rm ph}$} & \colhead{$T_0$\tablenotemark{b}} & \colhead{$M_{^{56}\rm Ni}$\tablenotemark{c}} \\ 
\colhead{} & \colhead{(days)} & \colhead{($10^{42}$erg/s)} & \colhead{(km/s)} & \colhead{(days)} & \colhead{(${\rm M}_{\odot}$)}}

\startdata
2014L & 13.0  & 2.06& 7650 & 90 & 0.075 \\
1994I & 12.0    & 2.63 &9900 & 85& 0.089 \\
2007gr & 14.0  & 1.72& 6700 & 125 & 0.066 
\enddata    
\tablecomments{Observed parameters of SNe Ic, 1994I, 2014L, and 2007gr. The typical errors for the measurement are 10-20\%. }
\tablenotetext{a}{Rise time from the shock breakout to the peak of quasi-bolometric curves in Figure \ref{<bolo>}.}
%\tablenotetext{b}{Photospheric velocity in unit of $10^9$ cm/s.}
\tablenotetext{b}{Derived from the light curves at $t > +50$ days.}
\tablenotetext{c}{Derived from equation \ref{eqNi} and the quasi-bolometric  curves in Figure \ref{<bolo>}.}

\label{Tab:Input}
\end{deluxetable}
%%%%%%%%%%%%%%

%%%%%%%%%%%%%%%%%

\section{Explosion Parameters}
\label{sect:Para}

Some basic explosion parameters, e.g.,  synthesized $^{56}$Ni mass ($M_{\rm Ni}$), ejecta mass ($M_{\rm{ej}}$), and ejecta kinetic energy  ($E_{\rm{ke}}$)  can be estimated from the observed peak flux ($L_{\rm max}$) and rise time ($t_r$)  of the bolometric light curve,  and the photospheric velocity ($v_{\rm ph}$) around the peak brightness with the assumption of a constant opacity.  For example,  $M_{\rm Ni}$ can be estimated using the Arnett law \citep{Arn82,Stritzinger05}:
\begin{equation}
M_{\rm Ni}=\frac{{L_{\rm max}}}{10^{43} \rm erg\ s^{-1}} \times (6.45 e^{-{t_r \over \tau_{\rm Ni}}}+1.45 e^{-{t_r \over \tau_{\rm Co}}})^{-1},
\label{eqNi}
\end{equation}

where, $\tau_{\rm Ni} = 8.8$ and $\tau_{\rm Co} = 111.3$ are the decay time of \Nifs\ and $^{56}$Co. 

Besides, following \citet{Arn82} and the simplifications in \B{WJC 15}, we write: 

\begin{mathletters}
\begin{displaymath}
%\begin{eqnarray}
M_{\rm ej}=\frac{1}{2}\frac{\beta c}{\kappa} v_{\rm ph} t^2_r ~~~~~~~~~~~~~~~~~~~~~~~~~~~~~~~~~~~~~~~~~~~~~~
\end{displaymath}

\begin{equation}
=0.77{\rm M}_{\odot}\left(\frac{\kappa}{0.1\rm cm^2g^{-1}}\right)^{-1}v_{\rm ph,9}\left(\frac{t_r}{10\rm d}\right)^2,
\label{eqMejp}
%\end{eqnarray}
\end{equation}

\end{mathletters}

\begin{mathletters}
\begin{displaymath}
~~E_{\rm ke}=\frac{3}{10}M_{\rm ej}v^2_{\rm ph} = \frac{3}{20}\frac{\beta c}{\kappa}v^3_{\rm ph}t^2_r~~~~~~~~~~~~~~~~~~~~~~~~~~~~~~~~~~~~~~~~
\end{displaymath}
\begin{equation}
%\begin{aligned}
~~~~~~=4.6\times10^{50} {\rm erg}\left(\frac{\kappa}{0.1 \rm{cm}^2g^{-1}}\right)^{-1}v_{\rm ph,9}^3 \left(\frac{t_r}{10d}\right)^2,
%\end{aligned}
\label{eqEkep}
\end{equation}
\end{mathletters}

where $\kappa$ is the effective optical opacity,  $v_{\rm ph,9}$ is the velocity at the photosphere in units of $10^9$ cm s$^{-1}$.

On the other hand,  $M_{\rm{ej}}$ and $E_{\rm{ke}}$ can be derived from the late-time tail of the light curves,
\begin{equation}
M_{\rm ej}=\frac{3}{10}\frac{v_{\rm ph}^2T_0^2}{C\kappa_{\gamma}}, ~~~~~~~~~~~~~~~~~~~~~~~~~~~~~~~~~~~~~~~~~~~~
\label{eqMetail}
\end{equation}

\begin{equation}
E_{\rm ke}=\left(\frac{3}{10}\right)^2 \frac{v^4_{\rm ph}T^2_0}{C\kappa_{\gamma}},~~~~~~~~~~~~~~~~~~~~~~~~~~~~~~~~~~~~~~
\label{eqEketail}
\end{equation}
 where $C$ is a dimensionless structure constant, typically about 0.05, $\kappa_{\gamma}$ is the gamma-ray opacity, $T_0$ is the gamma-ray trapping time scale (as defined in \citealt{Clocchiatti97}) and can be derived from the observed later-time light curves (see the details in \B{WJC 15}). 

Explosion parameters of  SN\,2014L, SN\,1994I, and SN\,2007gr  are measured and compared at below because  they show many similarities in observation.

\subsection{Rise time}
The rise time $t_r$ is essential to infer the mass of synthesized \Nifs\ via equation \ref{eqNi}. It can be measured from the very early light curve. 

\begin{figure}
\centering
\includegraphics[width=8cm,angle=0]{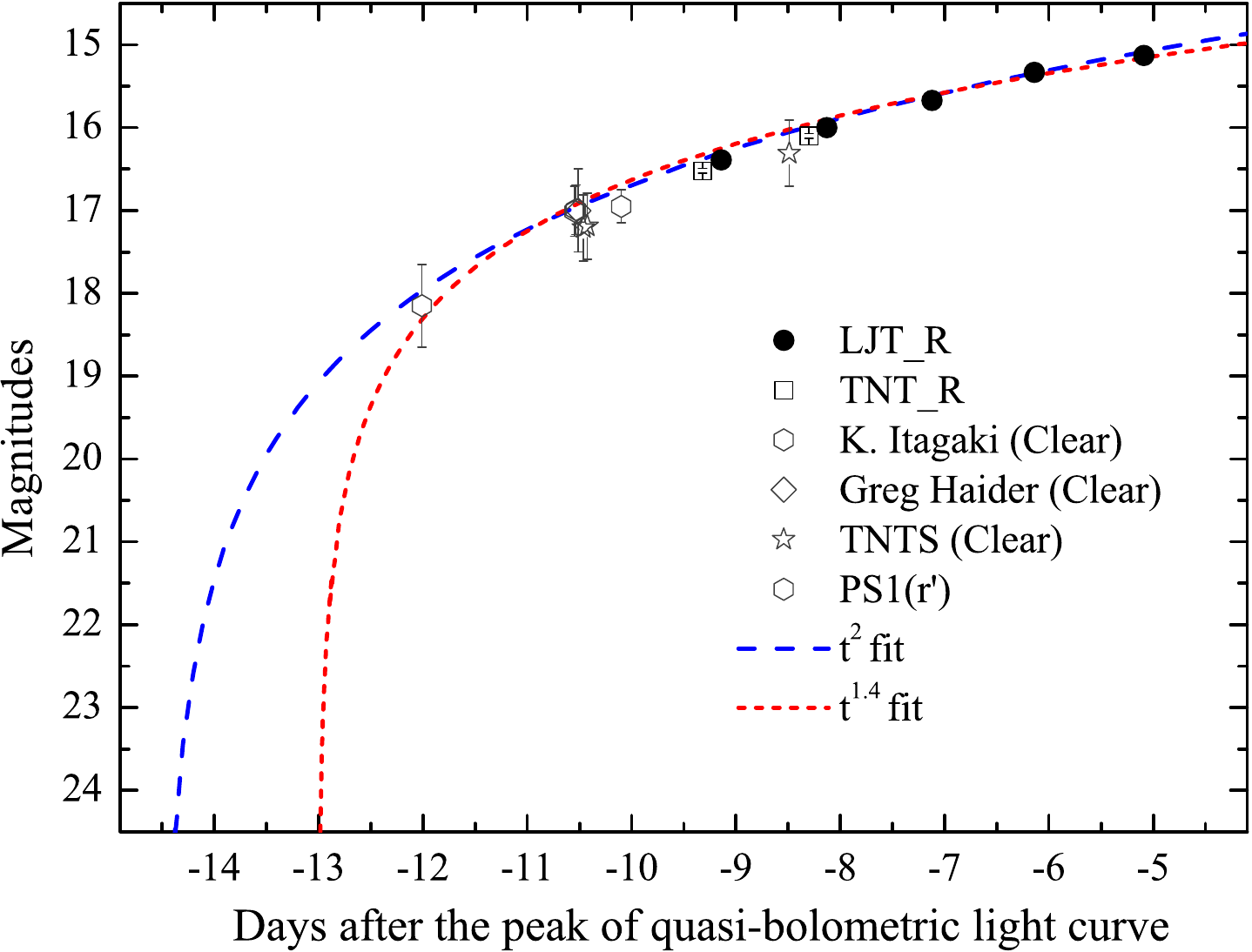}
 \caption{$t^n$ fitting for the early observation of SN\,2014L. The exponent of different curves are marked.}
\label{<rise>}
\end{figure}

Three groups reported the detection of SN\,2014L in their pre-discovery images after the discovery report of \citet{TMZhang14}.  The first detection was obtained by Koichi Itagaki (KI) through a 0.5 m reflector telescope at  the Takamizawa station on January 24.85 \citep{Yamaoka14} in a clear band image (roughly $R = 18.15 \pm 0.50$ mag).  Later on, Pan-STARRS1 reported the detection  for this SN, marked as PS1-14kd,  with the $r=17.01$ mag on UT 2014 January 26. 49. Gregory Haider  also detected it on January 26.54 in the clear images  (roughly $R = 17.00\pm0.40$ mag)  using the  12.5 inches reflector telescope at Christmas Valley. Combining these pre-discovery detections  with the early $R$-band light curve obtained by LJT and TNT,  the shock breakout date of SN\,2014L was derived on 2014 January 22.5 (MJD= 56679.5 $\pm$ 1.0) through the $t^2$ law approximation of the fireball model, see Figure \ref{<rise>}.  It suggests that SN\,2014L reached its bolometric maximum at about 14.5 days after the shock breakout. 

However, the fireball model might not be valid in the case of SNe Ic. The physics behind this model is that the increasing internal heating roughly compensates the adiabatic cooling of the ejecta due to Ni-decay. Since the envelope of SNe Ic is shock-heated, the Ni-decay plays less  of a role than it does in the case of SNe Ia (e.g., SNe Ia mag have 5-10 times more nickel than SNe Ic).  Thus, the fireball model may not be capable of giving an  accurate prediction for the moment of shock breakout. Furthermore, some poor-fit results of the fireball model in the case of very-young SNe Ia would suggest a steeper power-law index (e.g., \citealp{Zheng13,Zheng14}). 

We noted that some previous works (e.g., \citealt{Valenti08,Chen14};  \B{WJC 15}) suggested that the rise time of SN\,2007gr is about 13-14 days.  Given the narrower light curve of SN\,2014L, as presented in Figure \ref{<LC_comp>}, its rise time might be shorter than that of the SN\,2007gr.  Moreover, the light-curve modeling in  section \ref{subsect:LCM} suggests $t_r=13$ days of SN\,2014L.  Meanwhile, we liberalized the exponent of the power law and found a value of 1.4 with a $t_r$=13 days.  Considering the light-curve comparison of SN\,2007gr, the following light curve modeling, and the result of $t^{1.4}$ fit, we adopted $t_r=13$ days as the rise time of SN\,2014L in the related calculations. The rise time suggests that the shock breakout time of this SN is on  2014 January 24.8  (MJD = 56682.8$\pm$1.0). It indicates that the first detection of SN\,2014L taken by KI is only a few hours after the shock breakout.  Thus, SN\,2014L might be the earliest detected SN Ic to date.

%%%%%%%%%%%%%%%%%%%%%%%
%%%%%%%%%%%%%%%%%%%%%%%

\subsection{Mass and energy of the ejecta}

The $t_r$, $L_{\rm max}$, $v_{\rm ph}$, and $T_0$ parameters of SN\,2014L, SN\,1994I, and SN\,2007gr derived from the light curve and spectra are listed in Table \ref{Tab:Input}.  A fiducial optical  opacity $\kappa=0.1 \rm {cm}^2 \rm{g}^{-1}$, and a fiducial gamma-ray opacity $\kappa_{\gamma}=0.03 \rm {cm}^2 \rm{g}^{-1}$ are taken to calculate  $M_{\rm ej}$ and $E_{\rm ke}$. In Table \ref{Tab:MejEke}, as noted by \B{WJC 15},  parameters derived from the peak do not match those arising from the tail. That relates to the fiducial value of $\kappa$ and $\kappa_{\gamma}$ adopted in the calculation. If we set the results derived from the peak  equal to that from the tail, a relation about $\kappa$ and $\kappa_{\gamma}$ can be written as

\begin{mathletters}
\begin{displaymath}
\kappa=\frac{5}{3}(\beta c)C\frac{\kappa_{\gamma}}{v_{\rm ph}}\left(\frac{t_r}{T_0}\right)^2  ~~~~~~~~~~~~~~~~~~~~~~~~~~~~~~~~
\end{displaymath}
\begin{equation}
=0.01 \rm{cm}^2\rm{g}^{-1}\frac{\kappa_{\gamma}/0.03 \rm{cm}^2\rm{g}^{-1}} {\it v_{\rm ph,9}}\left(\frac{\it {t}_{r, \rm 10}}{\it {T}_{\rm 0,100}}\right).~~~~
\label{eqka}
\end{equation}
\end{mathletters}
\B{WJC 15} assumed that $\kappa_{\gamma}$\ is less dependent on explosion parameters than $\kappa$. Thus, $\kappa$ derived from equation \ref{eqka}, where $\kappa_{\gamma}=0.03 \rm {cm}^2 \rm{g}^{-1}$, is close to 0.02-0.03 ${\rm cm}^2 \rm{g}^{-1}$. This is much smaller than the assumed $\kappa=0.1 \rm {cm}^2 \rm{g}^{-1}$.
\begin{deluxetable}{lcccccr}

\tablecaption{Mass and Kinetic Energy of Ejecta}

\tablehead{\colhead{Parameters} & \colhead{14L$_{\rm p}$\tablenotemark{a}} & \colhead{94I$_{\rm p}$\tablenotemark{a}}& \colhead{07gr$_{\rm p}\tablenotemark{a}$ }  & \colhead{14L$_{\rm t}$\tablenotemark{b}}  & \colhead{94I$_{\rm t}$\tablenotemark{b} }  & \colhead{07gr$_{\rm t}$\tablenotemark{b} } }
\startdata
%$R_0(10^{11}$cm)	&	4.00	&	4.00	&	4.00	\\
$M_{\rm{ej}}$(${\rm M}_{\odot}$) &	1.00	&1.10  &1.01   &3.32	 & 5.49   &5.26 \\
$E_{\rm{ke}}$(foe)                       &	0.35	&0.75  &0.23   &1.16 &3.77    &	1.41 
\enddata    
\tablecomments{$M_{\rm{ej}}$ and $E_{\rm{ke}}$ of SN\,2014L, SN\,1994I, and SN\,2007gr. The typical errors for the measurement are 15-25\%.}
\tablenotetext{a}{Derived from the light-curve peaks based on equations \ref{eqMejp} and \ref{eqEkep} with $\kappa=0.10 \rm{cm}^2 \rm{g}^{-1}$.}
\tablenotetext{b}{Derived from the late-time tails based on equations \ref{eqMetail} and \ref{eqEketail} with $\kappa_{\gamma}=0.03 \rm{cm}^2 \rm{g}^{-1}$.}

\label{Tab:MejEke}
\end{deluxetable}

\subsection{Light-Curve Modeling}
\label{subsect:LCM}

\begin{deluxetable}{lccr}

\tablecaption{Parameters of Light-curve model}

\tablehead{\colhead{Parameters} & \colhead{Model A\tablenotemark{a}} & \colhead{Model B\tablenotemark{b}} & \colhead{Model C\tablenotemark{c}} }
\startdata
 \cutinhead{ Parameters for SN\,2014L}
%$R_0(10^{11}$cm)\tablenotemark{d}	&	4.00	&	4.00	&	4.00	\\
$M_{\rm{ej}}$(${\rm M}_{\odot}$)\tablenotemark{d} 	&	1.00	&	3.32	&	0.83	\\
$E_{\rm{ke}}$(foe)\tablenotemark{d} 	&	0.35	&	1.16	&	0.80	\\
$E_{\rm{th}}$(foe)\tablenotemark{d} 	&	1.10	&	1.10	&	0.10	\\
$E_{\rm{ke}}/E_{\rm{th}}$	&	0.32	&	1.05	&	8.00	\\
$\kappa$(cm$^2$/g)\tablenotemark{d}	&	0.10	&	0.03	&	0.12	\\
$T_0$(d)\tablenotemark{e}	&	55.1	&	100.5	&	30.3	\\
$T_+$(d)\tablenotemark{e}	&	871.3	&	1588.9	&	478.3	\\
$v_{\rm ph}$(km/s)\tablenotemark{e}	&	7659	&	7652	&	12710	\\
 \cutinhead{ Parameters for SN\,1994I}							
%$R_0(10^{11}$cm)\tablenotemark{d}	&	4.00	&	4.00	&	4.00	\\
$M_{\rm{ej}}$(${\rm M}_{\odot}$)\tablenotemark{d} 	&	1.10	&	5.49	&	0.83	\\
$E_{\rm{ke}}$(foe)\tablenotemark{d} 	&	0.75	&	5.17	&	1.30	\\
$E_{\rm{th}}$(foe)\tablenotemark{d} 	&	1.20	&	0.70	&	0.10	\\
$E_{\rm{ke}}/E_{\rm{th}}$	&	0.63	&	7.39	&	13.00	\\
$\kappa$(cm$^2$/g)\tablenotemark{d}	&	0.10	&	0.02	&	0.14	\\
$T_0$(d)\tablenotemark{e}	&	41.4	&	78.7	&	23.7	\\
$T_+$(d)\tablenotemark{e}	&	654.7	&	1244.6	&	375.2	\\
$v_{\rm ph}$(km/s)\tablenotemark{e}	&	10690	&	12563	&	16202	\\
 \cutinhead{ Parameters for SN\,2007gr}							
%$R_0(10^{11}$cm)\tablenotemark{d}	&	4.00	&	4.00	&	4.00	\\
$M_{\rm{ej}}$(${\rm M}_{\odot}$)\tablenotemark{d} 	&	1.01	&	5.06	&	0.78	\\
$E_{\rm{ke}}$(foe)\tablenotemark{d} 	&	0.23	&	1.17	&	0.50	\\
$E_{\rm{th}}$(foe)\tablenotemark{d} 	&	0.86	&	0.85	&	0.10	\\
$E_{\rm{ke}}/E_{\rm{th}}$	&	0.27	&	1.38	&	5.00	\\
$\kappa$(cm$^2$/g)\tablenotemark{d}	&	0.10	&	0.02	&	0.13	\\
$T_0$(d)\tablenotemark{e}	&	68.7	&	152.5	&	36.0	\\
$T_+$(d)\tablenotemark{e}	&	1085.5	&	2411.3	&	568.6	\\
$v_{\rm ph}$(km/s)\tablenotemark{e}	&	6178	&	6225	&	10365	

\enddata    
\tablecomments{Model parameters for the synthetic light curves of SN\,2014L, SN\,1994I, and SN\,2007gr. The typical errors for the measurement are 15-25\%.}
\tablenotetext{a}{Derived from the peak light curves.}
\tablenotetext{b}{Derived from the tail light curves.}
\tablenotetext{c}{Best fitting for the light curve at the first full month after shock break.}
\tablenotetext{d}{Input parameters of LC2.2 code.}
\tablenotetext{e}{Output parameters of LC2.2 code.}
\label{Tab:Output14L}
\end{deluxetable}
%%%%%%%%%%%%%%

%%%%%%%%%%%%%%%%%%%%%%%
\begin{figure}
\centering
\includegraphics[width=8.5cm,angle=0]{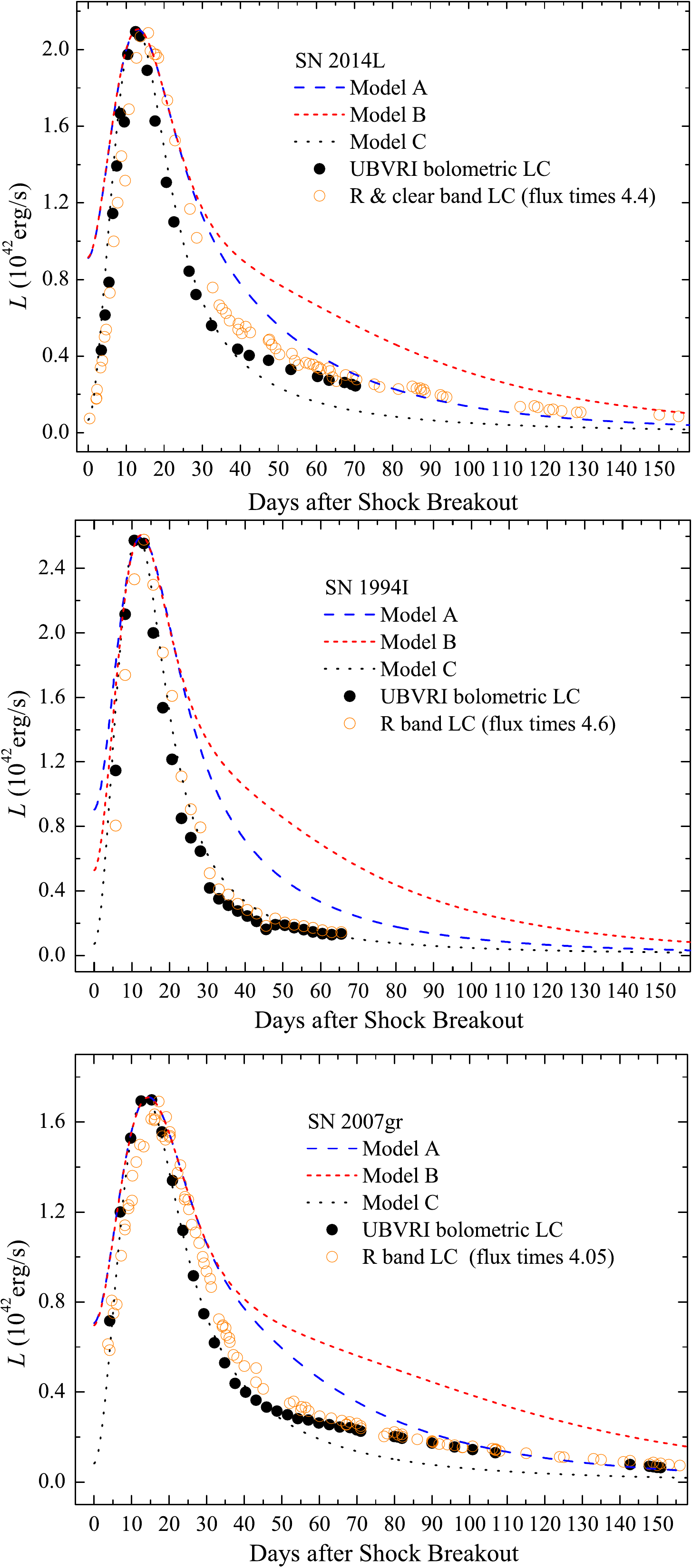}
 \caption{Modeling fitting for the bolometric light curve of SN\,2014L, SN\,1994I, and SN\,2007gr. Only the core of the LC2 model was adopted. The main input parameters are listed in Table \ref{Tab:Output14L}, see details in the text. }
\label{<LC2>}
\end{figure}
%%%%%%%%%%%%%%%%%%%%%%%
The bolometric light curve of SN\,2014L, SN\,1994I, and SN\,2007gr have been modeled by 
applying the LC2.2 code\footnote{\tt http://titan.physx.u-szeged.hu/$\sim$nagyandi/LC2.2} to check the explosion parameters estimated at above. This code is the most recent version of the LC2 code presented by \citet{Nagy16}.
LC2.2 computes the bolometric light curve of a homologous expanding supernova
using the radiative diffusion approximation as introduced by \citet{arnettfu}. The list
of model parameters includes the initial radius ($R_0$), the ejecta mass ($M_{\rm ej}$),
the initial \Nifs\ mass ($M_{\rm Ni}$), the total kinetic and thermal energy of the ejecta
($E_{\rm ke}$ and $E_{\rm th}$, respectively), the type and the exponent of the density
profile (power-law or exponential), and the average value of the Thompson-scattering
opacity ($\kappa$) in the envelope. 

Recombination of hydrogen and/or helium, which is built-in the code, is not relevant in the case of the 
Type Ic SNe. Thus, it has been turned off, similar to the additional energy input from magnetar spin-down.
LC2.2 assumes fixed gamma-ray and positron opacities,
set to their nominal values of $\kappa_{\gamma} = 0.027$ and $\kappa_{+} = 7$ cm$^2$g$^{-1}$, respectively. Thus, they are not adjustable parameters, unlike in the previous version described by \citet{Nagy16}.
Note also that while LC2.2, in principle, can model a two-component
ejecta having different masses, radii, density profiles, and opacities, we applied only a single component fit to the light curve of these SNe Ic.

The initial radius is not a sensitive parameter in this code. Thus, a constant $R_0=4\times10^{11}$cm is adopted for the following calculations. Table \ref{Tab:Output14L} lists the input parameters of SN\,2014L, SN\,1994I, and SN\,2007gr.  $M_{\rm Ni}$ is derived from equation \ref{eqNi} based on the optical flux because the $UBVRI$ bolometric light curves are modeled here.  $M_{\rm ej}$ and $E_{\rm ke}$ in Model A and B are derived from the peak (i.e., equation \ref{eqMejp} and \ref{eqEkep}) and tail  (i.e., equation \ref{eqMetail} and \ref{eqEketail}) of the light curves, respectively.  A fiducial optical  opacity $\kappa=0.1 \rm {cm}^2 \rm{g}^{-1}$ is adopted in Model A. $E_{\rm th}$ is a free parameter to match the peak between observation and calculation.

Figure \ref{<LC2>} displays the modeling results based on the input parameters in Table \ref{Tab:Output14L}. The $U$- and $B$-band flux are decreased much  more quickly than that in the $VRI$-bands. Thus, the full $UBVRI$-band photometry  only covers the first 70 days of SN\,2014L. Besides, it is reasonable to assume that the $R$-band light curve is a good proxy of the bolometric light curve within a constant scaling factor (\B{WJC 15}).  For comparison, we over-plotted the scaled R-band light curve by assuming the R-band luminosity scales with the bolometric luminosity. 

These two models yield a broader light curve than the observation. We noted that the width of light curve produced by LC2.2 code is inversely proportional to the scale of $\kappa$ and  the ratio of $E_{ke}$ and $E_{th}$. Thus, we  liberalized the parameters $\kappa$, $E_{ke}$ and $E_{th}$,  to find a  better fit for the observation. In this case, $M_{ej}$ follows $\kappa$ based on the relation of equation \ref{eqMejp}.

The best fitting recorded and plotted as Model C, matches the observations well at $\tau\lesssim40-70$ days.  For example, Model C gives a good fit for SN\,2014L and SN\,2007gr at about $\tau\lesssim40-50$ days, and the latter curves of the model are fainter than the data. This might be due to the leakage of  gamma-rays.  The model C of SN\,1994I gives a good fit during all of the observational epochs. It is uncertain after $\tau>70$ days because of the lack of observation. However, the tendency of the observed light curve is to follow the prediction of model C. In brief, model C yields a good fit at $\tau\lesssim40-70$ days, and model A gives a good fit for the tail at $\tau>90$ days. The mismatch between these models is probably due to the oversimplifying assumption of constant opacity.  As mentioned before, the rise time derived from model C (i.e., $t_r$=13 days ) is adopted in the above calculation because this model gives proper fitting during the rising phase. 

It is remarkable that the ratio of $E_{\rm ke}$/$E_{\rm th}$ in model C is much bigger than that of model A and B. Besides, SN\,1994I shows the narrowest light curve in this figure, and it has the highest $E_{\rm ke}$/$E_{\rm th}$ ratio. It seems that the narrower light curve could be the consequence of the  faster speed of ejecta.  However, this is vigorously challenged by BL SNe Ic, which shows the highest velocity and the broadest light curve. On the other hand, the speeds derived from model C are much higher than  those
observed, which indicates an overestimation of $E_{\rm ke}$ for these SNe.

\section{Summary}
\label{sect:con}

We have presented densely sampled photometric and spectroscopic observations of the Type Ic SN\,2014L. This SN was likely detected at only a few hours after shock breakout and reached a peak brightness of $L_{\rm max}=(2.06\pm0.50)\times10^{42}\rm{erg s}^{-1}$ at  $\tau\sim$ 13 days. The synthesized $^{56}$Ni derived from the peak bolometric brightness is $M(\rm Ni) = 0.075\pm 0.025$\Msun.  It is noted that some explosion parameters of SN\,2014L obtained from the peak brightness mismatch those derived from the tail of the light curve. We leave the issue of the peak/tail conflict for future studies, and only give  a range for these parameters for SN\,2014L as $M_{\rm{ej}}=(1.00-3.32){\rm M}_{\odot}$, $E_{\rm{ke}}=(0.35-1.16)$foe, $E_{\rm {th}}=1.10$ foe. Nevertheless, the results derived from the light curve around the peak are adopted in the literature more commonly.

A simple light-curve code was utilized to fit the bolometric light curve, which yields some proper fittings at some individual epochs.  The mismatched intervals between observation and modeling might be related to the more complicated behavior of optical opacity of ejecta. This code assumes spherical ejecta, which may not be valid in all SNe Ic. The spherical ejecta assumption may also contribute to the failure of fitting the whole LC, in addition to the issues with the constant $\kappa$. 

Besides, from the morphological comparison, SN\,2014L looks more-or-less similar to SN\,2007gr and SN\,1994I in both photometry and spectroscopy.  The differences, however, are also remarkable, e.g., the width of the light curve, the strength of \CaII\ NIR triplet, and the velocity of ejecta. SN\,2014L shows a stronger \CaII\ NIR triplet than SN\,2007gr, SN\,1994I, and even the Ca-rich events (e.g., SN\,2005E, SN\,2012hn) during the early phase. However, the ratio of the strengths of O/Ca and Si/Ca  in SN\,2014L is larger than the Ca-rich events, and it is close to the typical SNe Ic. Thus, the strong \CaII\ NIR triplet of SN\,2014L might suggest the enrichment of IMEs in  the outer layer of ejecta.  The light curve width, peak brightness, ejecta velocity, ejecta mass, synthesized $^{56}$Ni, and explosive energy of SN\,2014L locate between SN\,2007gr and SN\,1994I.  Therefore, the observation of SN\,2014L strengthens the physical relation between SN\,1994I and SN\,2007gr.

\acknowledgments
We thank the anonymous referee very much for his/her constructive suggestions which helped to improve the paper a lot. We acknowledge the support of the staff of the Li-Jiang 2.4 m telescope (LJT) and Xing-Long 0.8 m telescope (TNT), 2.16m telescope (XLT). Funding for the LJT has been provided by the Chinese Academy of Sciences (CAS) and the People's Government of Yunnan Province. The LJT is jointly operated and administrated by Yunnan Observatories and Center for Astronomical Mega-Science, CAS.  We also thanks Koichi Itagaki, Greg Haider, and David Bishop for providing the pre-discovery images about this SN.  J. Zhang is supported by the National Natural Science Foundation of China (NSFC, grants 11403096, 11773067); X. Wang is supported by NSFC (grants 11178003, 11325313, and 11633002) and the Major State Basic Research Development Program (2013CB834903);  J. Vink\'{o} is supported by GINOP-2.3.2-15-2016-00033 grant of the National Research, Development and Innovation Office (NKFIH) Hungary, funded by the European Union; J. Bai is supported by the NSFC (grants 11133006, 11361140347), the Strategic Priority Research Program ``The Emergence of Cosmological Structures" of the CAS (grant No. XDB09000000), and the Key Research Program of the CAS (Grant NO. KJZD-EW-M06);  L. Chang  is supported by the NSFC (grants 11573069); Y. Yang is supported through a Benoziyo Prize Postdoctoral Fellowship; J. Mao is supported by the NSFC (grants 11673062) and Hundred Talent Program and the Overseas Talent Program of Yunnan Province;  Y. Zhang  is supported by the NSFC (grants 11661161016); J. Wang is supported by the NSFC (grants 11303085); Y. Xin  is supported by the NSFC (grants 11573067). This work is also supported by the Western Light Youth Project; the Youth Innovation Promotion Association of the CAS (grants 2018081, 2015043, 2018080, 2016054); the Key Laboratory for Research in Galaxies and Cosmology of the CAS; and the Open Project Program of the Key Laboratory of Optical Astronomy, NAOC, CAS.

\clearpage

\end{document}